# A Capability Maturity Model for Urban Dataset Meta-data

| Mark S Fox | Bart Gajderowicz | Dishu Lyu |
|---|---|---|
| *Urban Data Centre* | *Urban Data Centre* | *Urban Data Centre* |
| *School of Cities* | *School of Cities* | *School of Cities* |
| *University of Toronto* | *University of Toronto* | *University of Toronto* |
| msf@mie.utoronto.ca | bartg@mie.utoronto.ca | lester.lyu@mail.utoronto.ca |

Draft version: October 11, 2024

**Abstract**

In the current environment of data generation and publication, there is an ever-growing number of datasets available for download. This growth precipitates an existing challenge: sourcing and integrating relevant datasets for analysis is becoming more complex. Despite efforts by open data platforms, obstacles remain, predominantly rooted in inadequate metadata, unsuitable data presentation, complications in pinpointing desired data, and data integration. This paper delves into the intricacies of dataset retrieval, emphasizing the pivotal role of metadata in aligning datasets with user queries. Through an exploration of existing literature, it underscores prevailing issues such as the identification of valuable metadata and the development of tools to maintain and annotate them effectively. The central contribution of this research is the proposition of a dataset metadata maturity model. Deriving inspiration from software engineering maturity models, this framework delineates a progression from rudimentary metadata documentation to advanced levels, aiding dataset creators in their documentation efforts. The model encompasses seven pivotal dimensions, spanning content to quality information, each stratified across five maturity levels to guide the optimal documentation of datasets, ensuring ease of discovery, relevance assessment, and comprehensive dataset understanding. This paper also incorporates the maturity model into a data cataloguing tool called CKAN through a custom plugin, CKANext-udc. The plugin introduces custom fields based on different maturity levels, allows for user interface customisation, and integrates with a graph database, converting catalogue data into a knowledge graph based on the Maturity Model ontology.

# Introduction

The world is awash in data, but we cannot seem to find the data we need. Though open data platforms strive to make it easier to find relevant data, Ojo et al. (2016) observe that common obstacles to finding relevant data are "poor metadata, failure to present data appropriately to different audiences and difficulty in locating data of interest." Metadata is defined in the Oxford English dictionary as "data whose purpose is to describe and give information about other data." Ojo et al. (2016) describes dataset metadata as referring to information such as "title, author, subjects, keywords, publisher, revision history, changes, and the source of data."

Kunze and Auer (2013) define the problem of dataset retrieval as determining the most relevant datasets according to a user query. This is part of the exploration phase of information seeking, which Koesten et al. (2017) define as determining "whether a search result, in our case a dataset, matches an information need." In understanding the process of exploration, it is important to distinguish between the direct examination of the contents of a dataset[1] versus the examination of its metadata. Chapman et al. (2020), in their survey of dataset search identify the following "open problems for metadata":

1. identifying the metadata that is of highest value to users with respect to datasets;
2. tools to automatically create and maintain that metadata; and
3. automatic annotation of dataset with metadata—linking them automatically to global ontologies

Research has explored each of these problems. Regarding the first problem, Datasheets for Datasets (Gebru et al., 2021) suggest over 50 metadata questions whose answers would enable a dataset consumer to determine the relevance of a dataset. Using Datasheets is a daunting task, requiring more information than the dataset cataloguer has available or is willing to provide. If the cataloguer is only willing to provide limited information, which is the most important to provide? A second problem is that the answers to the Datasheets questions are textual. That is, they are in a natural language such as English, which must be read by the dataset consumer. Existing keyword-based search may help narrow down the target datasets, but with possibly poor precision and recall.

In order to achieve higher levels of precision and recall, applied ontology methods have been used to construct vocabularies and ontologies to represent dataset metadata. The most prominent is the Data Catalog Vocabulary - DCAT (Albertoni et al., 2020) and schema.org. Other relevant vocabularies/ontologies include PROV (Lebo et al., 2013) for provenance and DQV (Albertoni & Isaac, 2016) for dataset quality.

Regarding the second and third problems, Neumaier et al. (2017) automate the harvesting of metadata descriptions of over 854K datasets from over 261 data portals and mapping them onto the DCAT. In addition, they add both provenance and quality metadata.

The difficulty in finding relevant datasets is exacerbated by the global growth of dataset repositories that would need to be searched. Open Data Portal Watch (Neumaier et al., 2017) collects metadata from more than 260 repositories, focusing on government data. Noy et al. (2019) have developed Google Dataset Search (GDS) to address this problem. GDS crawls the web to extract the metadata from various types of datasets, which is then transformed into a searchable knowledge graph.

The problem addressed in this paper is how does the cataloguer of a dataset know what types of metadata should be defined and selected from the myriad of dataset metadata entities and attributes? The dataset cataloguer may not be versed in the many relevant vocabularies, such as DCAT, schema.org, PROV and DQV. Datasheets specify over 50 questions requiring a considerable amount of detailed information. The DCAT vocabulary specifies 13 classes with over 70 properties, not including extensions for provenance, quality, etc. Consequently, dataset cataloguers face a bewildering number of metadata entities and attributes to use to document their datasets.

[1] Which falls into the area *sensemaking* (Russell et al., 1993).

In order to navigate the morass of potential entities and attributes that can be used to catalogue a dataset, our goal is to define a dataset metadata capability maturity model that identifies and sequences the metadata that should be provided by a dataset cataloguer. Our capability maturity model specifies a sequence of levels of dataset cataloguing capability starting with the base case, e.g., "no capability", followed by levels of increasing capability. It is analogous to the Capability Maturity Model for Software that offers software organizations guidance on controlling their development and maintenance processes and advancing toward a culture of software engineering and management excellence (Paulk et al., 1993). Each step up the maturity "ladder" specifies a more enhanced process capability. In the context of dataset metadata, the lowest level (base case) is having no metadata specification capability. The question then is what to include in the next and subsequent capability maturity levels such that we balance the amount of effort required to catalogue a dataset with sufficient metadata to discover, determine the suitability/relevance of a dataset and provide sufficient information on who and how it may be used.

Our methodology for determining the dataset metadata capability maturity model (DMCMM) is to review the literature on dataset search behaviours to ascertain both the types of information searched for and the frequency of its search. Our proxy for the latter is how often a type of information appears in the search literature. We then review existing vocabularies relevant to the representation of dataset metadata. To demonstrate the application of this work, we present use cases on how metadata properties of the maturity model can be used to determine the application of various data sharing frameworks, such as FAIR (Findable, Accessible, Interoperable, Reusable) and OCAP[2] (Ownership, Control, Access, Possession) for indigenous data (Mecredy et al., 2018).

# Dataset Metadata Requirements

Requirements for dataset metadata have emerged from a variety of sources. One source is based on the analysis of dataset search behaviours. Another source is an emerging, implicit consensus across the growing number of data platforms as to the metadata attributes should be provided. A third source is the metadata required to evaluate conformance to FAIR principles. A fourth reflects requirements specific to Indigenous data. This section reviews many of these requirements.

## Requirements Based on Search Behaviour

How can the search for relevant datasets be supported by the DMCMM? In this section we review the literature on dataset searching to understand what metadata is used to search for datasets, and their frequency of use, in order to determine the level in the DMCMM they should appear.

Koesten et al. (2017) studied information-seeking behaviour. The following are examples of search:

> "someone trying to find the number of schools in a given post code area would need to extract the answer from a larger dataset containing all entries for all schools in a country in 2016. Someone studying how the number of schools across different regions has changed over time would need to process and aggregate several versions of the same data, published year after year. Finally, school data could be mixed with

[2] https://fnigc.ca/ocap-training/

house prices statistics to understand how one aspect influences the other."

Koesten et al. proposed a five-pillar model for how people seek information (Table 1). They also identified three categories of metadata to describe a dataset: Relevance, Usability and Quality. Table 2 summarizes the metadata of a dataset they found important.

**Table 1.** Information Seeking Stages.

| Task → | Search → | Evaluate → | Explore → | Use |
|---|---|---|---|---|
| goal or process oriented<br>linking<br>time series analysis<br>summarising<br>presenting<br>exporting | STRATEGIES:<br>web search<br>engines data portals<br>asking people<br>FOI request | relevance[1]<br>usability [2]<br>quality[3] | basic visual scan<br>obvious errors<br>summarizing statistics<br>headers<br>documentation<br>metadata | |

**Table 2.** Categories of Metadata.

| Assess | Information needed about |
|---|---|
| Relevance | context, coverage, original purpose, granularity, summary, time frame |
| Usability | labeling, documentation, license, access, machine readability, language used, format, schema, ability to share |
| Quality | collection methods, provenance, consistency of formatting / labeling, completeness, what has been excluded |

Kacprzak et al. (2019) conducted an extensive analysis of web portal search logs and written requests. Their findings identified a dataset's topic area as important along with geospatial, temporal and dataset format properties, with geospatial and temporal at varying levels of granularity. An examination of the requests reproduced in their paper indicates the central role of domain expertise in the terms used in queries. These requests contain detailed descriptions of the topics where the terms are drawn from the domain's vocabulary.

Chen et al. (2019) analysed almost 2,000 dataset queries from online communities Stack Overflow, Open Data Stack Exchange, Reddit, and data.gov.uk. Table 3 depicts the results of their analysis. It divides the analysis into two categories, Metadata, where the queries refer to information contained in metadata, and Content, where the queries refer to information contained in the content of the dataset. Note that 94% of the queries refer to the domain or topic of the dataset, 50% refer to concepts and their properties in the dataset, 20% refer to geospatial information, 16% refer to the format of the dataset, and 10% to temporal information.

**Table 3.** Analysis of Dataset Queries.

| Category | Title | % of Queries | Example Query |
|---|---|---|---|
| Metadata | Name | 3.54% | HUST-ASL Dataset |
| | Domain/Topic | 94.45% | weather dataset with solar radiance and solar energy production |
| | Data Format | 16.23% | jpg images for all Unicode characters |
| | Language | 3.90% | annotated movie review dataset in German |
| | Accessibility | 7.40% | open source handwritten English alphabets dataset |
| | Provenance | 0.21% | FDA datasets about medicine name and the result has adverse events |
| | Statistics | 2.98% | dataset contains at least 1000 examples of opinion articles |
| | Overall | 96.05% | |
| Content | Concept | 50.59% | dataset about people, include gender, ethnicity, name |
| | Geospatial | 19.21% | judicial decisions in France |
| | Other Entities | 0.41% | datasets with nutrition data for many commercial food products (i.e., Lucky Charms, Monster Energy, Nutella, etc.) |
| | Temporal | 9.35% | 2011-2013 MoT failure rates on passenger cars |
| | Other Numbers | 1.59% | businesses that employ over 1000 people in Yorkshire region |
| | Overall | 63.79% | |

Kacprzak et al. (2019), in their analysis of query logs from four national open data platforms, identified the following metadata attributes as relevant:

- Geospatial
- Temporal
- Topic taxonomy
- Price
- License
- Format
- Size

Chua et al. (2020) analysed the information-seeking behaviours of 21 people using open data portals. Spatial and temporal keywords dominated the search queries and were supplemented with format and source filters. Follow up interviews identified that dataset incompleteness and outdatedness as issues.

Sharifpour et al. (2022) in their analysis of search behaviours based on different levels of domain expertise, discovered that expert users used more words and succeeded with shorter sessions, confirming one of the results of White et al. (2009). They also observed that dataset search is more difficult due to "the data for relevance judgement is not readily accessible within the metadata of datasets".

## Dataset Platform Metadata Requirements

The second source of requirements stem from the growing number of dataset platforms that are operating around the world. We determine these requirements by reviewing the literature on the metadata attributes that are found in dataset platforms. These platforms represent a growing consensus of the attributes deemed to be needed to support both search and accessibility.

Assaf et al. (2015) proposed the Harmonized Data modeL (HDL) that adopts and extends key properties of DCAT, Schema.org, CKAN, etc, to "ensure complete metadata coverage to enable data discovery, exploration and reuse." Their analysis identifies eight information types to be encoded as metadata:

1. General Information such as title and description.
2. Access Information such as the URL and license.
3. Ownership information such as author and maintainer.
4. Provenance information such as creation date and versioning.
5. Geospatial information such as geographic coverage.
6. Temporal information such as temporal span and granularity.
7. Statistical information property distribution and number of entities.
8. Quality information such as the quality of the data and metadata

Neumaiier et al. (2017) identified the following new or custom metadata properties (Table 4) in their analysis of over 749K CKAN datasets (referred to as "extra keys" in CKAN). They also add quality (DQV) (Albertoni & Isaac, 2016) and provenance (PROV) (Lebo et al., 2013) information to the dataset's metadata.

**Table 4.** Extra Keys.

| Extra key | Portals | Datasets | Mapping |
|---|---|---|---|
| spatial | 29 | 315,652 | dct:spatial |
| harvest_object.id | 29 | 514,489 | ? |
| harvest_source.id | 28 | 486,388 | ? |
| harvest_source_title | 28 | 486,287 | ? |
| guid | 21 | 276,144 | dct:identifier |
| contact-email | 17 | 272,208 | dcat:contactPoint |
| spatial-reference-system | 16 | 263,012 | ? |
| metadata-date | 15 | 265,373 | dct:issued |

The DataCite project (Rueda et al., 2017) seeks to create an interoperable e-infrastructure for research data. It highlights the importance of unique, persistent identifiers in datasets for achieving an interoperable e-infrastructure. "Persistent identifiers allow different platforms to exchange information consistently and unambiguously and provide a reliable way to track citations and reuse." In addition is the adoption of a common set of metadata properties, partitioned into mandatory, recommended and optional (Table 5).

**Table 5.** DataCite Metadata Properties.

| Mandatory | Recommended | Optional |
|---|---|---|
| Identifier | Subject | Language |

| Mandatory | Recommended | Optional |
|---|---|---|
| Creator | Contributor | Alternate Identifier |
| Title | Date | Size |
| Publisher | Related Identifier | Format |
| Publication Year | Description | Version |
| Resource Type | Geolocation | Rights |

Fenner et al. (2019) define a roadmap for data citation. They identify two types of metadata that need to be represented. This first is citation metadata. Table 6 lists the types of citation metadata in the first column and the corresponding properties as found in Dublin Core, Schema.org, DataCite and DATS (Sansone et al., 2017).

**Table 6.** Citation Metadata.

| Citation Metadata | Dublin Core | Schema.org | DataCite | DATS |
|---|---|---|---|---|
| Dataset Identifier | identifier | @id | identifier | identifier |
| Title | title | name | title | title |
| Creator | creator | author | creator | creator |
| Data repository or archive | publisher | publisher | publisher | publisher |
| Publication Date | date | datePublished | publication Year | date |
| Version | not available | version | version | version |
| Type | type | type | resourceTypeGeneral | type |

The second is discovery metadata used to enable the discovery of relevant datasets (Table 7).

**Table 7.** Discovery Metadata.

| Discovery Metadata | Dublin Core | Schema.org | DataCite | DATS |
|---|---|---|---|---|
| Description | description | description | description | datatype is a dimension, isAbout Material Material |
| Keywords | subject | keywords | subject | keywords |
| License | license | license | rights | license |
| Related Dataset | isPartOf is VersionOf references | isPartOf citation | relatedIdentifier | isPartOf |
| Related Publication | bibliographicCitation | citation | relatedIdentifier | publication |

Chapman et al. (2020) state that repositories need to consider data provenance, annotations, quality, granularity of content, data schema, language, and temporal coverage.

Thornton et al. (2021) analysed several Canadian open health data repositories regarding the richness of their metadata. As part of their analysis they used metadata defined in the Dataverse North metadata best practices guide (Cooper et al., 2019) and Data Citation Roadmap (Fenner et al., 2019). The following are the metadata in the Dataverse North guide:

- Title
- Author
- Description
- Subject
- Producer
- Contact Name
- Contact Affiliation
- Contact Email

Gebru et al. (2021) in their "Datasheets for Datasets" proposal defined 56 questions to document the provenance of machine learning datasets. These questions are divided into 7 categories:

1. Motivation: Who created the dataset? For what purpose? Who funded it?
2. Composition: What is the dataset composed of? Size? Completeness?
3. Collection Process: How was the data collected? When? Ethical process?
4. Preprocessing/Cleaning/Labeling: Was any cleaning or labeling performed?
5. Uses: How has the data been used? What can it be used for, or not?
6. Distribution: How and when will the dataset be distributed? Any restrictions?
7. Maintenance: Who supports the dataset? Will it be updated? Will older versions be maintained?

Appendix 2 contains the complete list of questions for each category.

## Licensing Metadata

Another important category of metadata are the licenses that dictate who and how a dataset may be used. To ascertain the metadata required for the latter, we review licenses under which datasets are often published.

The Creative Commons Organization has six types of licenses[3] that span the continuum from free use of the material for both commercial and non-commercial uses, to limitations on remixing, adapting, and building upon, and for commercial use. Across all these licenses is the requirement to give attribution to the creator of the material.

The Open Knowledge Foundation has three types of licenses[4] that focus specifically on data. The licenses allow users of the data to:

- share: To copy, distribute and use the database.
- create: To produce works from the database.
- adapt: To modify, transform and build upon the database.

[3] https://creativecommons.org/about/cclicenses/
[4] https://opendatacommons.org/licenses/

Similar to the Creative Commons license, attribution is required (for two of the licenses) for any public use of the data and its derivations. In both cases knowing the creator or owner and the license is important.

## FAIR

As FAIR principles continue to grow in adoption, the DMCMM needs to include attributes that support the FAIR evaluation of a dataset. Bahim et al. (2020) define a FAIR (Findable, Accessible, Interoperable, Reusable) Data Maturity Model. "The principles emphasise machine-actionability (i.e., the capacity of computational systems to find, access, interoperate, and reuse data with none or minimal human intervention) because humans increasingly rely on computational support to deal with data as a result of the increase in volume, complexity, and creation speed of data."[5]

A set of indicators have been defined to evaluate the "FAIRness" of a dataset. The indicators are divided into Essential, Important and Useful**Error! Reference source not found.**, and **Table 8**, **Table 9**, and **Table 10** list the indicators for each partition.

**Table 8.** Essential FAIR Indicators.

| FAIR | ID | Indicator |
|---|---|---|
| F1 | RDA-F1-01M | Metadata is identified by a persistent identifier |
| F1 | RDA-F1-01D | Data is identified by a persistent identifier |
| F1 | RDA-F1-02M | Metadata is identified by a globally unique identifier |
| F1 | RDA-F1-02D | Data is identified by a globally unique identifier |
| F2 | RDA-F2-01M | Rich metadata is provided to allow discovery |
| F3 | RDA-F3-01M | Metadata includes the identifier for the data |
| F4 | RDA-F4-01M | Metadata is offered in such a way that it can be harvested and indexed |
| A1 | RDA-A1-02M | Metadata can be accessed manually (i.e., with human intervention) |
| A1 | RDA-A1-02D | Data can be accessed manually (i.e., with human intervention) |
| A1 | RDA-A1-03M | Metadata identifier resolves to a metadata record |
| A1 | RDA-A1-03D | Data identifier resolves to a digital object |
| A1 | RDA-A1-04M | Metadata is accessed through standardised protocol |
| A1 | RDA-A1-04D | Data is accessible through standardised protocol |
| A1.1 | RDA-A1.1-01M | Metadata is accessible through a free access protocol |
| A2 | RDA-A2-01M | Metadata is guaranteed to remain available after data is no longer available |
| R1 | RDA-R1-01M | Plurality of accurate and relevant attributes are provided to allow reuse |
| R1.1 | RDA-R1.1-01M | Metadata includes information about the licence under which the data can be reused |
| R1.3 | RDA-R1.3-01M | Metadata complies with a community standard |
| R1.3 | RDA-R1.3-01D | Data complies with a community standard |

[5] https://www.go-fair.org/fair-principles/

| FAIR | ID | Indicator |
| --- | --- | --- |
| R1.3 | RDA-R1.3-02M | Metadata is expressed in compliance with a machine-understandable community standard |

**Table 9.** Important FAIR Indicators.

| FAIR | ID | Indicator |
| --- | --- | --- |
| A1 | RDA-A1-01M | Metadata contains information to enable the user to get access to the data |
| A1 | RDA-A1-05D | Data can be accessed automatically (i.e. by a computer program) |
| A1.1 | RDA-A1.1-01D | Data is accessible through a free access protocol |
| I1 | RDA-I1-01M | Metadata uses knowledge representation expressed in standardised format |
| I1 | RDA-I1-01D | Data uses knowledge representation expressed in standardised format |
| I1 | RDA-I1-02M | Metadata uses machine-understandable knowledge representation |
| I1 | RDA-I1-02D | Data uses machine-understandable knowledge representation |
| I2 | RDA-I2-01M | Metadata uses FAIR-compliant vocabularies |
| I3 | RDA-I3-01M | Metadata includes references to other metadata |
| I3 | RDA-I3-03M | Metadata includes qualified references to other metadata |
| R1.1 | RDA-R1.1-02M | Metadata refers to a standard reuse licence |
| R1.1 | RDA-R1.1-03M | Metadata refers to a machine-understandable reuse licence |
| R1.2 | RDA-R1.2-01M | Metadata includes provenance information according to community-specific standards |
| R1.3 | RDA-R1.3-02D | Data is expressed in compliance with a machine-understandable community standard |

**Table 10.** Useful FAIR Indicators.

| FAIR | ID | Indicator |
| --- | --- | --- |
| A1.2 | RDA-A1.2-01D | Data is accessible through an access protocol that supports authentication and authorisation |
| I2 | RDA-I2-01D | Data uses FAIR-compliant vocabularies |
| I3 | RDA-I3-01D | Data includes references to other data |
| I3 | RDA-I3-02M | Metadata includes references to other data |
| I3 | RDA-I3-02D | Data includes qualified references to other data |
| I3 | RDA-I3-04M | Metadata include qualified references to other data |
| R1.2 | RDA-R1.2-02M | Metadata includes provenance information according to a cross-community language |

### Indigenous Data Requirements

Metadata requirements for datasets containing indigenous data, stem from Indigenous Data Sovereignty, which addresses "the rights and interests of Indigenous Peoples in relation to data about them, their territories, and their ways of life" (Carroll et al., 2020). Several frameworks have been proposed including Canada's OCAP[6] (Mecredy et al., 2018) and Australia's guidance for indigenous data (Commonwealth of Australia, 2024). OCAP, developed by the First Nations Information Governance Centre[7], is a set of principles "regarding the collection, use and disclosure of data or information regarding first nations." It focuses on protecting indigenous individual privacy rights as well as the collective rights of community. OCAP is an acronym for Ownership, Control, Access and Possession[8]. Each are defined in the following excerpts from the OCAP training module:

- **Ownership:** "The notion of ownership refers to the relationship of a First Nations community to its cultural knowledge/ data/ information. The principle states that a community or group owns information collectively in the same way that an individual owns their personal information. Ownership is distinct from stewardship. The stewardship or custodianship of data or information by an institution that is accountable to the group is a mechanism through which ownership may be maintained. This can be done with data-sharing agreements and other legal instruments."

- **Control:** "The aspirations and inherent rights of First Nations to maintain and regain control of all aspects of their lives and institutions extend to information and data. The principle of 'control' asserts that First Nations people, their communities and their representative bodies must control how information about them is collected, used and disclosed. The element of control extends to all aspects of information management, from collection of data to the use, disclosure, and ultimate destruction of data."

- **Access:** First Nations must have access to information and data about themselves and their communities, regardless of where it is held. The principle also refers to the right of First Nations communities and organizations to manage and make decisions regarding who can access their collective information."

- **Possession:** "While 'ownership' identifies the relationship between a people and their data, possession reflects the state of stewardship of data. First Nations possession puts data within First Nations jurisdiction and therefore, within First Nations control. Possession is the mechanism to assert and protect ownership and control. First Nations generally exercise little or no control over data that is in possession of others, particularly other governments."

The guidance provided for the management of First Nations data in Australia contains guidance on how to work with indigenous communities along with elements of OCAP and FAIR (Commonwealth of Australia, 2024).

# Dataset Metadata Vocabularies

With the introduction of open data portals such as CKAN and Dataverse, interest in vocabularies for representing dataset metadata has grown. This section reviews vocabularies that have been developed in order to: 1) understand what metadata attributes the vocabularies have chosen to include, and 2) the terms they use for potential reuse in DMCMM.

[6] https://fnigc.ca/ocap-training/
[7] https://fnigc.ca/
[8] Reproduced from Module 1 of OCAP online training participant notes, developed by Algonquin College and FNIGC.

## VoID

Vocabulary of Interlinked Datasets[9] (Alexander et al., 2011) is one of the early RDF vocabularies for dataset metadata. It identifies Dublin Core Metadata terms to be used for datasets. Table 11 depicts the metadata terms. The prefix “dct” denotes the namespace “http://purl.org/dc/terms/”.

**Table 11.** Dublin Core Metadata terms for dataset metadata.

| Term | Purpose |
|---|---|
| dct:title | The name of the dataset. |
| dct:description | A textual description of the dataset. |
| dcterms:creater | An entity such as person organisation or service that is primarily responsible for creating the dataset. The creator should be described as an RDF resource, rather than just providing the name as literal. |
| dct:publisher | An entity such as a person organisation or service that is responsible for making the dataset available. The publisher should be described as an RDF resource, rather than just providing the name as a literal. |
| dct:contributor | An entity such as a person organisation or service that is responsible for making contributions to the dataset. The contributor should be described as an RDF resource, rather than just providing the name as a literal. |
| dct:source | A related resource from which the dataset is derived. The source should be described as an RDF resource rather than as literal. |
| dct:date | A point or period of time associated with an event in the life-cycle of the resource The value should be formatted and data-typed as an xsd:date. |
| dct:created | Date of creation of the dataset. The value should be formatted and data-typed as an xsd:date. |
| dct:issued | Date of formal issuance (e.g., publication) of the dataset. The value should be formatted and data-typed as an xsd:date. |
| dct:modified | Date on which the dataset was changed. The value should be formatted and data-typed as an xsd:date. |

In addition, it provides properties for contact information, licensing, dataset domain categories, format, access information and statistics. Table 12 lists the statistics-related properties. The prefix “void” denotes the namespace “http://rdfs.org/ns/void#”.

**Table 12.** VoID dataset statistics.

| Property | Purpose |
|---|---|
| void:triples | The total number of triples contained in the dataset. |

[9] https://www.w3.org/TR/void/

| Property | Purpose |
| --- | --- |
| void:entities | The total number of entities that are described in the dataset. To be an entity in a dataset, a resource must have a URI, and the URI must match the dataset's void:uriRegexPattern, if any. Authors of VoID files may impose arbitrary additional requirements, for example, they may consider any foaf:Document resources as not being entities. |
| void:classes | The total number of distinct classes in the dataset. In other words, the number of distinct class URIs occurring as objects of rdf:type triples in the dataset. |
| void:properties | The total number of distinct properties in the dataset. In other words, the number of distinct property URIs that occur in the predicate position of triples in the dataset. |
| void:distinctSubjects | The total number of distinct subjects in the dataset. In other words, the number of distinct URIs or blank nodes that occur in the subject position of triples in the dataset. |
| void:distinctObjects | The total number of distinct objects in the dataset. In other words, the number of distinct URIs, blank nodes, or literals that occur in the object position of triples in the dataset. |
| void:documents | If the dataset is published as a set of individual documents, such as RDF/XML documents or RDFa-annotated web pages, then this property indicates the total number of such documents. Non-RDF documents, such as web pages in HTML or images, are usually not included in this count. This property is intended for datasets where the total number of triples or entities is hard to determine. void:triples or void:entities should be preferred where practical. |

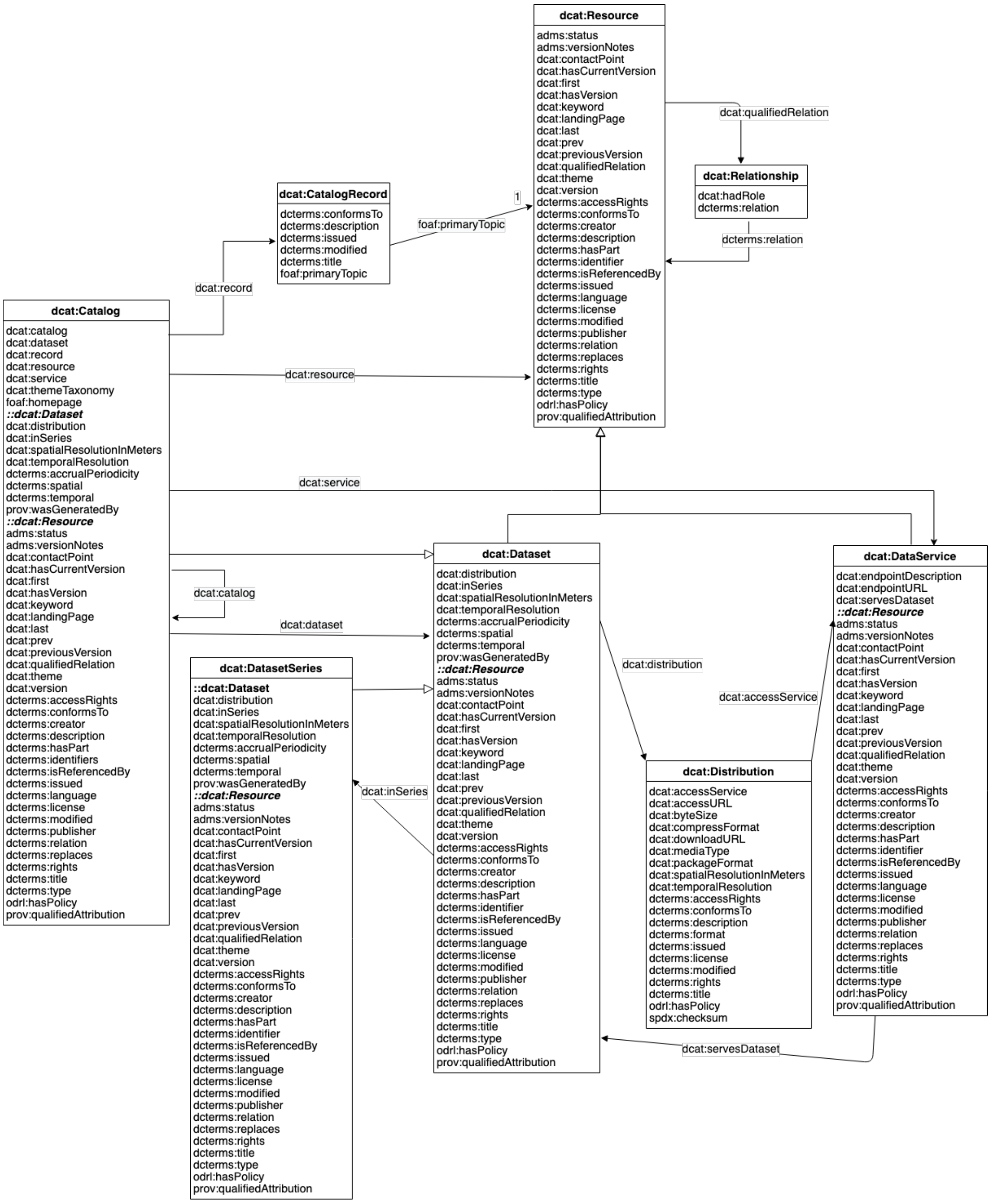

**Figure 1.** DCAT3 Classes and Properties (Albertoni et al., 2023).

## DCAT

The Data Catalog Vocabulary (DCAT)[10] is a W3C RDF-based vocabulary that enables interoperability among data catalogues published on the Web. The vocabulary defines a set of metadata terms for describing data catalogues and datasets. Figure 1 depicts the classes and properties in the version 3 (working draft) of DCAT. The dcat:Catalog class is used to define a web accessible catalogue composed of dcat:Resources of which dcat:Dataset and dcat:DataService are subclasses. A rich set of properties are provided to describe both. The prefix "dcat" denotes the namespace "http://www.w3.org/ns/dcat#".

[10] https://www.w3.org/TR/vocab-dcat-3/

## DCAT-AP

The DCAT-AP (Van Nueffelen, 2021) provides a standard for the description of dataset metadata, which is published by data portals across Europe. It identifies a required set of DCAT classes and properties and categorizes them as mandatory, recommended, and optional, which can be interpreted as a three-level maturity model. Table 13,

Table **14**, and Table 15 define the properties of the three categories, respectively.

**Table 13.** DCAT-AP Mandatory Dataset Properties.

| Property | URI | Range | Usage note | Card |
|---|---|---|---|---|
| Description | dct:description | rdfs:Literal | This property contains a free-text account of the Dataset. This property can be repeated for parallel language versions of the description. | 1.. n |
| Title | dct:title | rdfs:Literal | This property contains a name given to the Dataset. This property can be repeated for parallel language versions of the name. | 1..n |

**Table 14.** DCAT-AP Recommended Dataset Properties.

| Property | URI | Range | Usage note | Card |
|---|---|---|---|---|
| contact point | dcat:contactPoint | vcard:Kind | This property contains contact information that can be used for sending comments about the Dataset. | 0..n |
| dataset distribution | dcat:distribution | dcat: Distribution | This property links the Dataset to an available Distribution. | 0..n |
| keyword/ tag | dcat:keyword | rdfs: Literal | This property contains a keyword or tag describing the Dataset. | 0..n |
| publisher | dct:publisher | foaf:Agent | This property refers to an entity (organisation) responsible for making the Dataset available. | 0..1 |
| spatial/ geographical coverage | dct:spatial | dct:Location | This property refers to a geographic region that is covered by the Dataset. | 0..n |

| Property | URI | Range | Usage note | Card |
|---|---|---|---|---|
| temporal coverage | dct:temporal | dct:PeriodOfTime | This property refers to a temporal period that the Dataset covers. | 0..n |
| theme/category | dcat:theme, subproperty of dct:subject | skos:Concept | This property refers to a category of the Dataset. A Dataset may be associated with multiple themes. | 0..n |

**Table 15.** DCAT-AP Optional Dataset Properties.

| Property | URI | Range | Usage note | Card |
|---|---|---|---|---|
| access rights | dct:accessRights | dct:RightsStatement | This property refers to information that indicates whether the Dataset is open data, has access restrictions or is not public. A controlled vocabulary with three members (:public, :restricted, :non-public) will be created and maintained by the Publications Office of the EU. | 0..1 |
| creator | dct:creator | foaf:Agent | This property refers to the entity primarily responsible for producing the dataset | 0..1 |
| conforms to | dct:conformsTo | dct:Standard | This property refers to an implementing rule or other specification. | 0..n |
| documentation | foaf:page | foaf: Document | This property refers to a page or document about this Dataset. | 0..n |
| frequency | dct:accrualPeriodicity | dct:Frequency | This property refers to the frequency at which the Dataset is updated. | 0..1 |
| has version | dct:hasVersion | dcat:Dataset | This property refers to a related Dataset that is a version, edition, or adaptation of the described Dataset. | 0..n |
| identifier | dct:identifier | rdfs:Literal | This property contains the main identifier for the Dataset, e.g. the URI or other unique identifier in the context of the Catalogue. | 0..n |

| Property | URI | Range | Usage note | Card |
|---|---|---|---|---|
| is referenced by | dct:isReferencedBy | rdfs:Resource | This property provides a link to a description of a relationship with another resource | 0..n |
| is version of | dct:isVersionOf | dcat:Dataset | This property refers to a related Dataset of which the described Dataset is a version, edition, or adaptation. | 0..n |
| landing page | dcat:landingPage | foaf:Document | This property refers to a web page that provides access to the Dataset, its Distributions and/or additional information. It is intended to point to a landing page at the original data provider, not to a page on a site of a third party, such as an aggregator. | 0..n |
| language | dct:language | dct:LinguisticSystem | This property refers to a language of the Dataset. This property can be repeated if there are multiple languages in the Dataset. | 0..n |
| other identifier | adms:identifier | adms:Identifier | This property refers to a secondary identifier of the Dataset, such as MAST/ADS[1]5, DataCite [1]6, DOI17, EZID18 or W3ID19. | 0..n |
| provenance | dct:provenance | dct:ProvenanceStatement | This property contains a statement about the lineage of a Dataset. | 0..n |
| qualified attribution | prov:qualifiedAttribution | prov:Attribution | This property refers to a link to an Agent having some form of responsibility for the resource | 0..n |
| qualified relation | dcat:qualifiedRelation | dcat: Relationship | This property is about a related resource, such as a publication, that references, cites, or otherwise points to the dataset. | 0..n |
| related resource | dct:relation | rdfs: Resource | This property refers to a related resource. | 0..n |
| release date | dct:issued | rdfs: Literal typed as xsd:date or xsd:dateTime | This property contains the date of formal issuance (e.g., publication) of the Dataset. | 0..1 |

| Property | URI | Range | Usage note | Card |
|---|---|---|---|---|
| sample | adms:sample | dcat:Distribution | This property refers to a sample distribution of the dataset | 0..n |
| source | dct:source | dcat:Dataset | This property refers to a related Dataset from which the described Dataset is derived. | 0..n |
| spatial resolution | dcat:spatialResolutionInMeters | xsd:decimal | This property refers to the minimum spatial separation resolvable in a dataset, measured in meters. | 0..n |
| temporal resolution | dcat:temporalResolution | xsd:duration | This property refers to the minimum time period resolvable in the dataset. | 0..n |
| Type | dct:type | skos:Concept | This property refers to the type of the Dataset. A controlled vocabulary for the values has not been established. | 0..1 |
| update/ modification date | dct:modified | rdfs:Literal typed as xsd:date or xsd:dateTime | This property contains the most recent date on which the Dataset was changed or modified. | 0..1 |
| version | owl:versionInfo | rdfs:Literal | This property contains a version number or other version designation of the Dataset. | 0..1 |
| version notes | adms:versionNotes | rdfs:Literal | This property contains a description of the differences between this version and a previous version of the Dataset. This property can be repeated for parallel language versions of the version notes. | 0..n |
| was generated by | prov:wasGeneratedBy | prov:Activity | This property refers to an activity that generated, or provides the business context for, the creation of the dataset. | 0..n |

## Schema.org

Schema.org contains a number of classes and properties relevant to documenting datasets. Google provides a guide[11] for developers to enable dataset discovery. It distinguishes between required schem.org properties[12]: A sample of the schema.org class definition sin given in Appendix A.

[11] https://developers.google.com/search/docs/appearance/structured-data/dataset

- name – A descriptive name of a dataset (e.g., “Snow depth in Northern Hemisphere”)
- description – A short summary describing a dataset.

And recommended schema.org properties:

- url – Location of a page describing the dataset.
- sameAs – Other URLs that can be used to access the dataset page. A link to a page that provides more information about the same dataset, usually in a different repository.
- version – The version number or identifier for this dataset (text or numeric).
- isAccessibleForFree – Boolean (true|false) specifying if the dataset is accessible for free.
- keywords – Keywords summarizing the dataset.
- identifier – An identifier for the dataset, such as a DOI. (text, URL, or PropertyValue).
- variableMeasured – What does the dataset measure? (e.g., temperature, pressure)

## DQV

Data Quality Vocabulary[13] is an extension of DCAT that focuses on “the quality of the data, how frequently is it updated, whether it accepts user corrections, persistence commitments etc.” (Albertoni & Isaac, 2016). The following lists the core classes. The prefix “dqv” denotes the namespace “http://www.w3.org/ns/dqv#”.

- dqv:QualityAnnotation represents feedback and quality certificates given about the dataset or its distribution.
- dct:Standard represents a standard the dataset or its distribution conforms to.
- dqv:QualityPolicy represents a policy or agreement that is chiefly governed by data quality concerns.
- dqv:QualityMeasurement represents a metric value providing quantitative or qualitative information about the dataset or distribution.
- prov:Entity represents an entity involved in the provenance of the dataset or distribution.

Figure 2 depicts the core classes and properties of DQV.

[12] https://github.com/ESIPFed/science-on-schema.org/blob/master/guides/Dataset.md
[13] https://www.w3.org/TR/vocab-dqv/

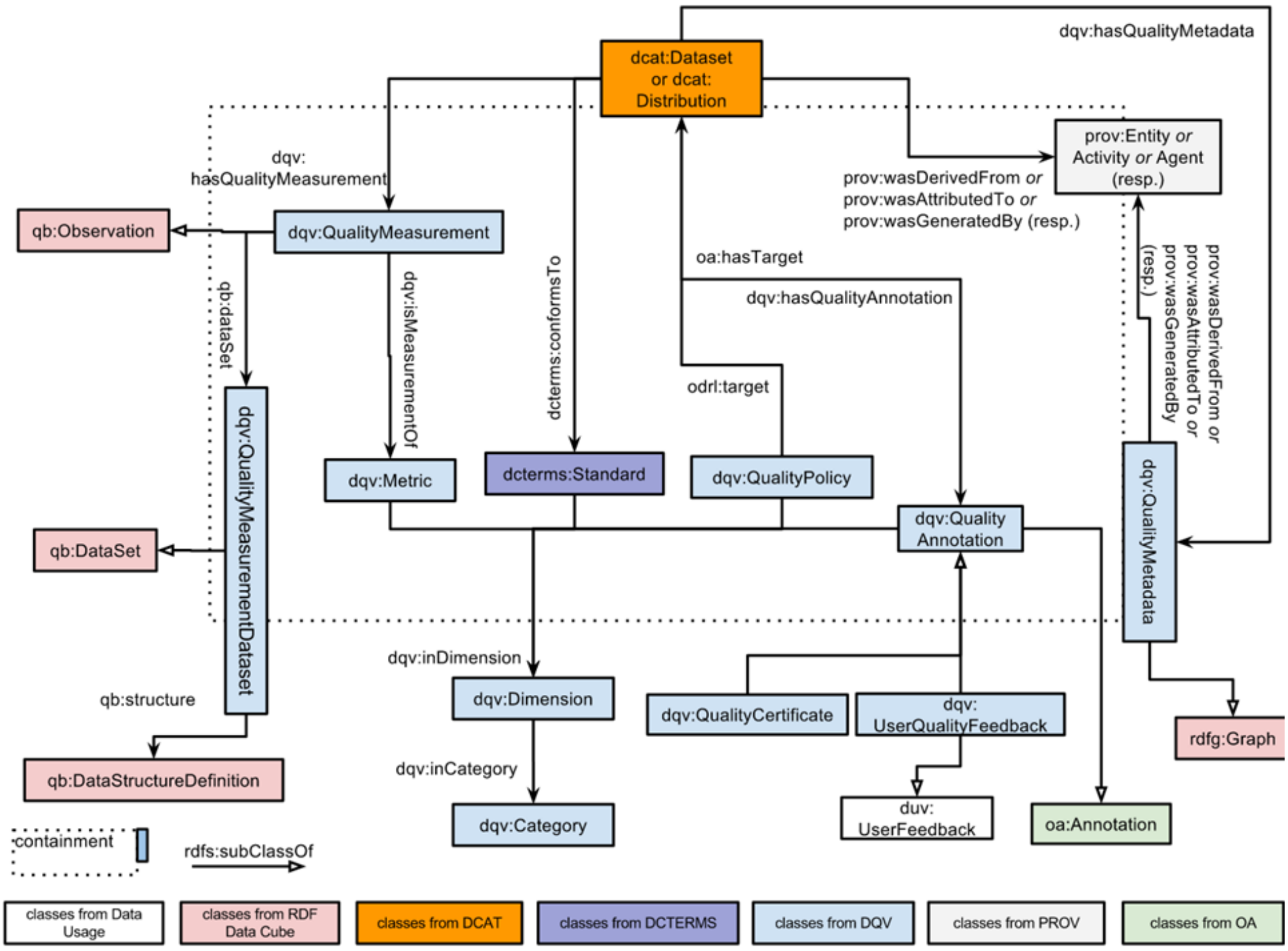

**Figure 2.** DQV Information Model (from Albertoni & Isaac, 2016)

## DDI

Data Documentation Initiative (DDI) has developed standards for documenting social sciences surveys and datasets (www.ddialliance.org). It provides a deeper dive into the properties describing the content of datasets and how they were generated (Thomas et al., 2014). Several dimensions of the content are described, including dataset provenance and analysis (DDI-Lifecycle) (Poynter and Spiegel, 2016), preservation and discovery (DDI-Codebook[14]), and a SKOS extension that includes statistical information about datasets and refinement of SKOS properties (XKOS) (Cotton et al, 2015). DDI metadata properties are viable for inclusion in the DMCMM, but as of this writing, DDI is not yet available as RDF or linked-data formats.

## ODRL

Open Digital Rights Language[15] (Iannella & Villata, 2018) "is a policy expression language that provides a flexible and interoperable information model, vocabulary, and encoding mechanisms for representing statements about the usage of content and services." It "represents Policies that express Permissions, Prohibitions and Duties related to the usage of Asset resources. The Information Model (Figure 3) explicitly expresses what is allowed and what is not allowed by the Policy, as well as other terms, requirements, and parties involved."

---

[14] https://ddi-alliance.atlassian.net/wiki/spaces/DDI4/pages/929792030/DDI+Codebook+Development+Work
[15] https://www.w3.org/TR/odrl-model/

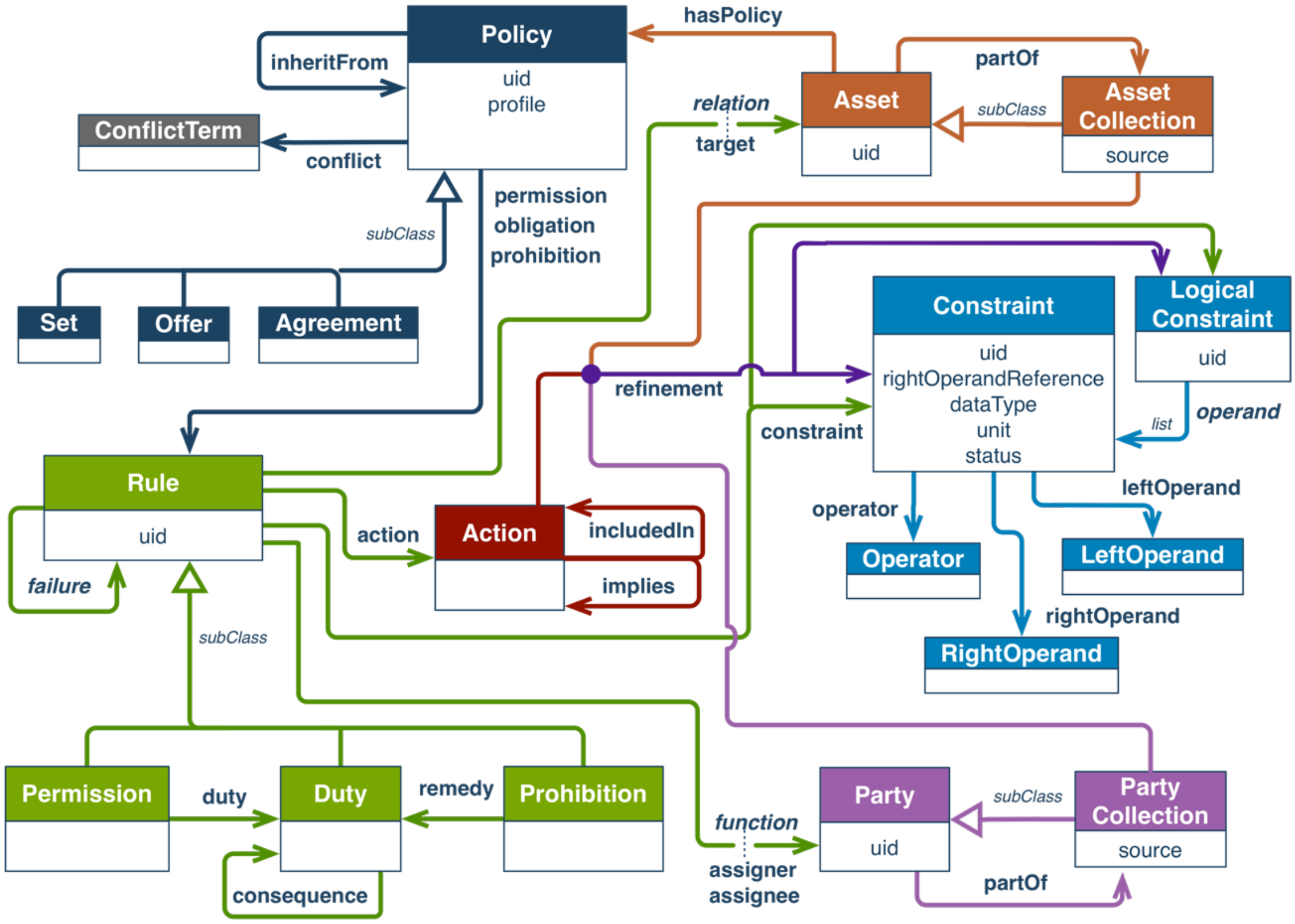

**Figure 3.** ODRL Information Model (from Iannella & Villata (2018)).

# A Capability Maturity Model for Dataset Metadata

As identified in the previous section, there is a plethora of information that a dataset producer could provide. But as observed by Gebru et al. (2021) not all is available or may take a significant effort to provide. Given the various backgrounds of dataset providers, and their depth of knowledge of the dataset they wish to catalogue, our goal is to simplify the process while assuring that the most important metadata is captured at the outset. Our method of simplifying the process is to define a maturity model that identifies and stratifies the metadata that should be provided by a dataset producer. By following the steps of the maturity model, the probability of acquiring the most important metadata from the dataset producer is increased. Secondly, by presenting the metadata requirements in a stratified format, it limits the perceived size and complexity of the task for the dataset producer.

Given the variety of information that can be used to document a dataset, we have chosen to first categorize the metadata properties by information categories. They are then combined into each maturity level based on the probability of acquiring most relevant metadata properties, as discussed above. Similar to Assaf et al. (2015), we define the following categories:

1. **Content** such as title and description
2. **Access** Information such as the URL and license.
3. **Ownership** information such as author and maintainer.
4. **Provenance** information such as creation date and versioning.
5. **Temporal/Geospatial** information such as geographic coverage and temporal span and granularity.
6. **Statistical** information property distribution and number of entities.
7. **Quality** information.

Where appropriate DCAT properties and classes are used for compatibility. Based on DCAT, a dataset is a dcat:Dataset object. A specific version of the dataset is a dct:Distribution. These are related by the property: dct:distribution (dctat:Dataset, dcat:Distribution). Depending on the domain of the property, the data resource being catalogued is either dct:Dataset or the dcat:Distribution related to the dct:Dataset.

The following prefixes are used in the model.

**Table 16.** Maturity Model Prefixes.

| Prefix | URI |
|---|---|
| adms | http://www.w3.org/ns/adms# |
| cc | http://creativecommons.org/ns# |
| cudr | http://data.urbandatacentre.ca/ |
| dc | http://purl.org/dc/elements/1.1/ |
| dcat | http://www.w3.org/ns/dcat# |
| dct | http://purl.org/dc/terms/ |
| dqv | http://www.w3.org/ns/dqv# |
| fair | http://ontology.eil.utoronto.ca/fair# |
| foaf | http://xmlns.com/foaf/0.1/ |
| oa | http://www.w3.org/ns/oa# |
| odrl | http://www.w3.org/ns/odrl/2/ |
| owl | http://www.w3.org/2002/07/owl# |
| prov | http://www.w3.org/ns/prov# |
| rdfs | http://www.w3.org/2000/01/rdf-schema# |
| sc | https://schema.org/ |
| skos | http://www.w3.org/2004/02/skos/core# |
| vann | http://purl.org/vocab/vann/ |
| vcard | http://www.w3.org/2006/vcard/ns# |
| void | http://rdfs.org/ns/void |
| xsd | http://www.w3.org/2001/XMLSchema# |

## Maturity Level 1

Maturity Level 1 focuses on findability. It is guided by our review of dataset search behaviours with the goal of making the dataset "findable" based on typical search behaviours. At level 1 we also want to limit the initial information provided by the cataloguer to that which is most often used in searching for datasets. Guided by the results of Chen et al. (2019), domain information, which in our model includes title, description and keywords is the most searched for information. Geospatial information, data format and temporal information were the next most searched for information. Others have included publication date. Note that DCAT-AP's mandatory attributes are a subset of level 1.

**Table 17.** Dataset Maturity Level 1: Findability.

| Category | Description | Property | Value Restriction |
|---|---|---|---|
| Content | Domain/Topic | dcat:theme | skos:Concept |
| | Title | dct:title | rdfs:Literal |
| | Description | dct:description | rdfs:Literal |
| | Keywords | dcat:keyword | rdfs:Literal |
| | Format (file type if relevant) | dct:format | dct:MediaType |
| | Dataset size in megabytes | datasetSize | xsd:integer |
| | Metadata identifier – to be used as a unique identifier for the catalogue entry | catalogueEntryIdentifier | rdfs:Literal |
| Provenance | Published date | dct:issued | xsd:datetime |
| Temporal/ Geospatial | Time period data spans | dct:temporal | dct:PeriodOfTime |
| | Geospatial area data spans | dct:spatial | dct:Location |

## Maturity Level 2

Maturity Level 2 focuses on characteristics of access including ownership. Once a dataset is "found", understanding who, how and where to access the dataset is the next most important metadata.

**Table 18.** Dataset Maturity Level 2: Access.

| Category | Description | Property | Value Restriction |
|---|---|---|---|
| Access | Access Category: Open, Closed, Service | accessCategory | {open, closed, service} |
| | License | dct:license | dct:LicenseDocument |
| | Limits on use (e.g., academic purposes – going beyond license) | odrl:hasPolicy | odrl:Policy |
| | Location of dataset: where it can be accessed | dcat:accessURL | rdfs:Resource |
| | Access service specification | dcat:accessService | dcat:DataService |
| | URL for a downloadable file | dcat:downloadURL | rdfs:Resource |
| Ownership | Owner | dct:rightsHolder | foaf:Agent |
| | Contact point | dcat:contactPoint | vcard:Kind |
| | Publisher | dct:publisher | foaf:Agent |

| Category | Description | Property | Value Restriction |
| --- | --- | --- | --- |
| | Creator | dct:creator | foaf:Agent |

Notes:

- **odrl:Policy** [16]: A non-empty group of Permissions (via the permission property) and/or Prohibitions (via the prohibition property) and/or Duties (via the obligation property). The Policy class is the parent class to the Set, Offer, and Agreement subclasses.
For odrl: properties, the data resource being catalogued is assumed to be an instance of the odrl:Asset class.

## Maturity Level 3

Maturity level 3 delves deeper into the content of the dataset. It covers identification, language and documentation, and whether the dataset contains synthetic data. It also supports the specification of whether the dataset contains information about individuals and indigenous data. Secondly, it expands on the Temporal/Geospatial aspects of the dataset

Figure 4: Dataset Maturity Level 3: Content

| Category | Description | Property | Value Restriction |
| --- | --- | --- | --- |
| Content | Unique identifier for the dataset. Often assigned by creator or publisher. Not necessarily persistent or globally unique. | dct:identifier | rdfs:Literal |
| | Language | dct:language | dct:LinguisticSystem |
| | Documentation | dcat:landingPage | foaf:Document |
| | Contains data about individuals | containsIndividualData | xsd:boolean |
| | Contains data about identifiable individuals | containsIdentifiableIndividualData | xsd:boolean |
| | Contains Indigenous Data | containsIndigenousData | xsd:boolean |
| | Contains synthetic data | containsSyntheticData | xsd:boolean |
| | Location of synthetic data generation documentation | syntheticDataDocumentation | rdfs:Resource |
| Temporal/ Geospatial | Temporal resolution | dcat:temporalResolution | xsd:duration |
| | Spatial resolution in meters | dcat:spatialResolutionInMeters | xsd:decimal |

[16] https://www.w3.org/TR/odrl-model/#policy

| Category | Description | Property | Value Restriction |
|---|---|---|---|
| | Spatial resolution by administrative area (e.g., a city or neighbourhood) | :spatialResolutionByAdminArea | sc:AdminstrativeArea |

## Maturity Level 4

Maturity Level 4 focuses primarily on provenance of a dataset, which includes versioning information, and linkages to other versions. Secondly, it expands on the Temporal/Geospatial aspects of the dataset.

**Table 19.** Dataset Maturity Level 4: Provenance

| Category | Description | Property | Value Restriction |
|---|---|---|---|
| Provenance | Version of the dataset | owl:versionInfo | rdfs:Literal |
| | Version notes | adms:versionNotes | rdfs:Literal |
| | Link to dataset that it is a version of | dct:isVersionOf | dcat:Dataset |
| | Link to datasets that are versions of it | dct:hasVersion | dcat:Dataset |
| | Provenance of the data | dct:provenance | dct:ProvenanceStatement |
| | Provenance document location | prov:wasQuotedFrom | prov:Entity |
| Temporal/ Geospatial | Temporal resolution | dcat:temporalResolution | xsd:duration |
| | Spatial resolution in meters | dcat:spatialResolutionIn Meters | xsd:decimal |
| | Spatial resolution by administrative area (e.g., a city or neighbourhood) | :spatialResolutionByAdm inArea | sc:AdminstrativeArea |

## Maturity Level 5

Maturity level 5 focuses on attributes relevant to indigenous data management policies. It includes properties relevant to the identification of whether the dataset contains data about communities that require additional policies that may have stricter privacy rules. For example, when an entire community owns a dataset, a steward acts as the rights holder on behalf of that community.

**Table 20.** Dataset Maturity Level 5: Indigenous.

| Category | Description | Property | Value Restriction |
|---|---|---|---|
| Access | Stewardship by an organization that is accountable to the community. | hasSteward | foaf:Organization |
| Ownership | Community permission (who gave permission) | communityRightsHolder | foaf:CommunityGroup |
| Temporal/ Geospatial | Communities from which data is derived | spatialCommunity (sub-property of dct:spatial) | dct:Location |

Notes:

- **odrl:Policy** [17]: A non-empty group of Permissions (via the permission property) and/or Prohibitions (via the prohibition property) and/or Duties (via the obligation property). The Policy class is the parent class to the Set, Offer, and Agreement subclasses.
  For odrl: properties, the data resource being catalogued is assumed to be an instance of the odrl:Asset class.
- **containsIdentifiableIndividualData**: Does the data hold identifiable individual data that can be used to uniquely identify an individual data was collected about? If yes, the dataset is not anonymized.

**Table 21.** CommunityGroup Class Definition.

| ClassProperty | Property | Value Restriction |
|---|---|---|
| CommunityGroup | rdfs:subClassOf | foaf:Group |

- **CommunityGroup**: A subclass of foaf:Group that is group of agents in a community.

- **Table 22.** communityRightsHolder Property Definition.

| ClassProperty | Property | Value Restriction |
|---|---|---|
| communityRightsHolder | rdfs:subPropertyOf | dct:rightsHolder |

- **communityRightsHolder**: An agent that has the rights to manage access rights to indigenous data. That person can be Indigenous themselves or a non-Indigenous agent that acts as the steward for access rights to the data.

**Table 23.** spatialCommunity Property Definition.

| ClassProperty | Property | Value Restriction |
|---|---|---|
| spatialCommunity | rdfs:subPropertyOf | dct:spatial |

[17] https://www.w3.org/TR/odrl-model/#policy

- **spatialCommunity**: A geospatial area occupied by or representative of a community.

## Maturity Level 6

Maturity Level 6 provides basic statistics and data quality. These are found less often in the literature but relevant in ascertaining relevance by the searcher. The attributes are derived from VOID and DQV.

**Table 24.** Dataset Maturity Level 6: Statistics and Quality.

| Category | Description | Property | Value Restriction |
|---|---|---|---|
| Statistical | If tabular dataset, number of rows | void:rows | xsd:positiveInteger |
| | If tabular dataset, number of columns | void:columns | xsd:positiveInteger |
| | If tabular dataset, the number of filled-in data cells | void:cells | xsd:positiveInteger |
| | If RDF dataset, total number of triples | void:triples | xsd:postiveInteger |
| | If RDF dataset, total number of entities in the dataset | void:classes | xsd:postiveInteger |
| | if RDF dataset, total number of properties in the dataset | void:properties | xsd:postiveInteger |
| Quality | Description of data quality. | dqv:hasQualityAnnotation | dqv:QualityAnnotation |
| | Metrics for data quality property, like completeness, accuracy, etc[18] | dqv:inDimension | dqv:Dimension |

Notes:

- **dvq:inDimension** : Represents the dimensions a quality metric, certificate and annotation allow a measurement of.
- **dqv:QualityAnnotation**[19]: Represents quality annotations, including ratings, quality certificates or feedback that can be associated to datasets or distributions. Quality annotations must have one oa:motivatedBy statement with an instance of oa:Motivation (and skos:Concept) that reflects a quality assessment purpose. This instance is defined as dqv:qualityAssessment.
- **dvq:Dimension**[20]: Represents criteria relevant for assessing quality. Each quality dimension must have one or more metric to measure it. A dimension is linked with a category using the dqv:inCategory property.

[18] https://www.w3.org/TR/vocab-dqv/#examples
[19] https://www.w3.org/TR/vocab-dqv/#dqv:QualityAnnotation
[20] https://www.w3.org/TR/vocab-dqv/#dqv:Dimension

# Capability Maturity Model Evaluation

In this section, the DMCMM metadata is evaluated as to its adequacy to support three uses: 1) dataset search; 2) FAIR evaluation; and 3) OCAP compliance.

## Search

Chen et al.'s (2019) analysis of search logs clearly identifies a stratification of metadata used to search for datasets, based on frequency. The topic of the dataset is the most used metadata for searching, far exceeding any other. Kacprzak et al. (2019) highlights the importance of a rich taxonomy of topics to aid the searcher. We note that this is also important in the cataloguing process. Topics are captured in level 1 by the dcat:theme property with a value restriction of skos:Concept. It is recommended that the platform being used to catalogue and search for datasets provide for taxonomies of topics across multiple domains.

Geospatial and temporal information about a dataset are the next highest terms used to search. Consequently, they are included in level 1 as dct:spatial with its values restricted to dct:PeriodOfTIme, and dct:temporal with its values restricted to dct:Location.

Data Format is the next highest metadata used in search. This is included in level 1 as dct:format with value restriction of dct:MediaType. Along with format, we included datasetSize property with a value of xsd:integer, as identified by Kacprzak et al.

In summary, from a search perspective, the most often used metadata attributes are included in level 1. Assuming that level 1 will be the most completed, by a cataloguer, basic search will be supported.

## FAIR

In order to support the FAIR evaluation of a dataset we identify for each FAIR indicator the attribute that can be used to evaluate the indicator. We assume that when metadata is referred to by FAIR it is referring to the attributes defined in DMCMM. When the maturity level is 0, this means that the DMCMM satisfies the indicator by definition. For example, whether rich metadata is provided for a dataset is by definition true if DMCMM is used as the metadata model. When the level is P, the answer is determined by the platform that the dataset is available. Our answer is provided for the platform described in the next section. It uses CKAN with a DMCMM plugin.

Some FAIR indicators do not have a corresponding DMCMM attributes but can be determined by examining either the metadata or dataset (Table 25 to Table 27). In this case we designate the level as E. Some DMCMM attributes may not directly address the indicator. For example, RDA-F1-01M asks if the dataset is identified by a persistent identifier. While the corresponding DMCMM attribute dct:identifier provides a unique identifier, it is not necessarily persistent. The software performing FAIR evaluation would have to examine the attribute's value to determine persistence.

Table 25. Essential FAIR properties and corresponding DMCMM properties.

| FAIR | ID | Essential Indicators | Level | DMCMM Attribute |
|---|---|---|---|---|
| F1 | RDA-F1-01M | Metadata is identified by a persistent identifier | 1 | catalogueEntryIdentifier |
| | RDA-F1-01D | Data is identified by a persistent identifier | 2 | dct:identifier |
| | RDA-F1-02M | Metadata is identified by a globally unique identifier | 1 | catalogueEntryIdentifier |

| FAIR | ID | Essential Indicators | Level | DMCMM Attribute |
|---|---|---|---|---|
| | RDA-F1-02D | Data is identified by a globally unique identifier | 2 | dct:identifier |
| F2 | RDA-F2-01M | Rich metadata is provided to allow discovery | 0 | DMCMM |
| F3 | RDA-F3-01M | Metadata includes the identifier for the data | 2 | dct:identifier |
| F4 | RDA-F4-01M | Metadata is offered in such a way that it can be harvested and indexed | P | CKAN+DMCMM |
| A1 | RDA-A1-02M | Metadata can be accessed manually (i.e., with human intervention) | P | CKAN+DMCMM |
| | RDA-A1-02D | Data can be accessed manually (i.e., with human intervention) | P | CKAN+DMCMM |
| | RDA-A1-03M | Metadata identifier resolves to a metadata record | P | CKAN+DMCMM |
| | RDA-A1-03D | Data identifier resolves to a digital object | E | dct:identifier |
| | RDA-A1-04M | Metadata is accessed through standardised protocol | P | CKAN+DMCMM |
| | RDA-A1-04D | Data is accessible through standardised protocol | 2 | dcat:accessService |
| A1.1 | RDA-A1.1-01M | Metadata is accessible through a free access protocol | P | CKAN+DMCMM |
| A2 | RDA-A2-01M | Metadata is guaranteed to remain available after data is no longer available | P | CKAN+DMCMM |
| R1 | RDA-R1-01M | Plurality of accurate and relevant attributes are provided to allow reuse | E | dcat:accessURL<br>dcat:accessService |
| R1.1 | RDA-R1.1-01M | Metadata includes information about the licence under which the data can be reused | 2 | dct:LicenseDocument |
| R1.3 | RDA-R1.3-01M | Metadata complies with a community standard | 0 | DMCMM is defined in terms of dc, dcat, etc. |
| R1.3 | RDA-R1.3-01D | Data complies with a community standard | E | dcat:accessURL<br>dcat:accessService |
| R1.3 | RDA-R1.3-02M | Metadata is expressed in compliance with a machine-understandable community standard | 0 | DMCMM is defined in terms of dc, dcat, etc. |

Table 26. Important FAIR properties and corresponding DMCMM properties.

| FAIR | ID | Important Indicators | Level | DMCMM Attribute |
|---|---|---|---|---|
| A1 | RDA-A1-01M | Metadata contains information to enable the user to get access to the data | 2<br>2<br>2<br>2<br>2 | accessCategory<br>dct:license<br>dcat:accessURL<br>dct:rightsHolder<br>dcat:contactPoint |

| FAIR | ID | Important Indicators | Level | DMCMM Attribute |
|---|---|---|---|---|
| | RDA-A1-05D | Data can be accessed automatically (i.e. by a computer program) | 2 | dcat:accessService |
| A1.1 | RDA-A1.1-01D | Data is accessible through a free access protocol | 2 | dcat:accessService |
| I1 | RDA-I1-01M | Metadata uses knowledge representation expressed in standardised format | 0 | DMCMM |
| | RDA-I1-01D | Data uses knowledge representation expressed in standardised format | E | |
| | RDA-I1-02M | Metadata uses machine-understandable knowledge representation | 0 | DMCMM |
| | RDA-I1-02D | Data uses machine-understandable knowledge representation | E | |
| I2 | RDA-I2-01M | Metadata uses FAIR-compliant vocabularies | 0 | DMCMM |
| I3 | RDA-I3-01M | Metadata includes references to other metadata | E | |
| | RDA-I3-03M | Metadata includes qualified references to other metadata | E | |
| R1.1 | RDA-R1.1-02M | Metadata refers to a standard reuse licence | 2 | dct:license |
| | | | 2 | accessCategory |
| | | | 5 | odrl:hasPolicy |
| | RDA-R1.1-03M | Metadata refers to a machine-understandable reuse licence | 2 | dct:license |
| R1.2 | RDA-R1.2-01M | Metadata includes provenance information according to community-specific standards | 4 | owl:versionInfo |
| | | | 4 | adms:versionNotes |
| | | | 4 | dct:isVersionOf |
| | | | 4 | dct:hasVersion |
| | | | 4 | dct:provenance |
| | | | 4 | prov:wasQuotedFrom |
| R1.3 | RDA-R1.3-02D | Data is expressed in compliance with a machine-understandable community standard | 2 | dcat:accessURL<br>dcat:accessService |

Table 27. Useful FAIR properties and corresponding DMCMM properties.

| FAIR | ID | Useful Indicators | Level | DMCMM Attribute |
|---|---|---|---|---|
| A1.2 | RDA-A1.2-01D | Data is accessible through an access protocol that supports authentication and authorisation | 2 | dcat:accessService |
| I2 | RDA-I2-01D | Data uses FAIR-compliant vocabularies | 2 | dcat:accessURL |
| I3 | RDA-I3-01D | Data includes references to other data | 2 | dcat:accessURL |
| | RDA-I3-02M | Metadata includes references to other data | 0 | Determined by evaluator |

| FAIR | ID | Useful Indicators | Level | DMCMM Attribute |
|---|---|---|---|---|
| | RDA-I3-02D | Data includes qualified references to other data | 2 | dcat:accessURL |
| | RDA-I3-04M | Metadata include qualified references to other data | 0 | Determined by evaluator |
| R1.2 | RDA-R1.2-02M | Metadata includes provenance information according to a cross-community language | 4 | owl:versionInfo |
| | | | 4 | adms:versionNotes |
| | | | 4 | dct:isVersionOf |
| | | | 4 | dct:hasVersion |
| | | | 4 | dct:provenance |
| | | | 4 | prov:wasQuotedFrom |

## OCAP

In Table 28, we identify possible indicators for each OCAP theme. For each indicator we identify the DMCMM attributes that can be used to evaluate the indicator.

Table 28. OCAP properties and corresponding DMCMM properties.

| | Indicator | Level | DMCMM Attribute |
|---|---|---|---|
| Ownership | Identify the Community that the data is drawn from. | 5 | spatialCommunity |
| | Identify the organization that "owns" the data. | 5 | communityRightsHolder |
| | Identify if the dataset contains indigenous data. | 3 | contansIndigenousData |
| Control | License to be agreed to by user. | 2 | dct:license |
| | Data sharing agreement that defines who, what and how data is to be shared – beyond the terms of the license. | 2 | odrl:hasPolicy |
| Access | Access methods and limitations. | 2 | accessCategory |
| | | 2 | dcat:accessURL |
| | | 1 | dct:format |
| | | 2 | dcat:downloadURL |
| Possession | Identify the steward that manages the data. | 5 | hasSteward |

# Implementation

The Maturity Model has been implemented as a CKAN[21] plugin (CKANext-udc[22]) to integrate the maturity model into the CKAN dataset cataloguing process. The plugin facilitates the inclusion of custom fields, allows for their reordering, and categorizes them into distinct maturity levels. It also allows for integration with a graph database to store each catalogue entry as a knowledge graph built on top of the ontology. The maturity model itself is defined as an ontology[23] and implemented in OWL.

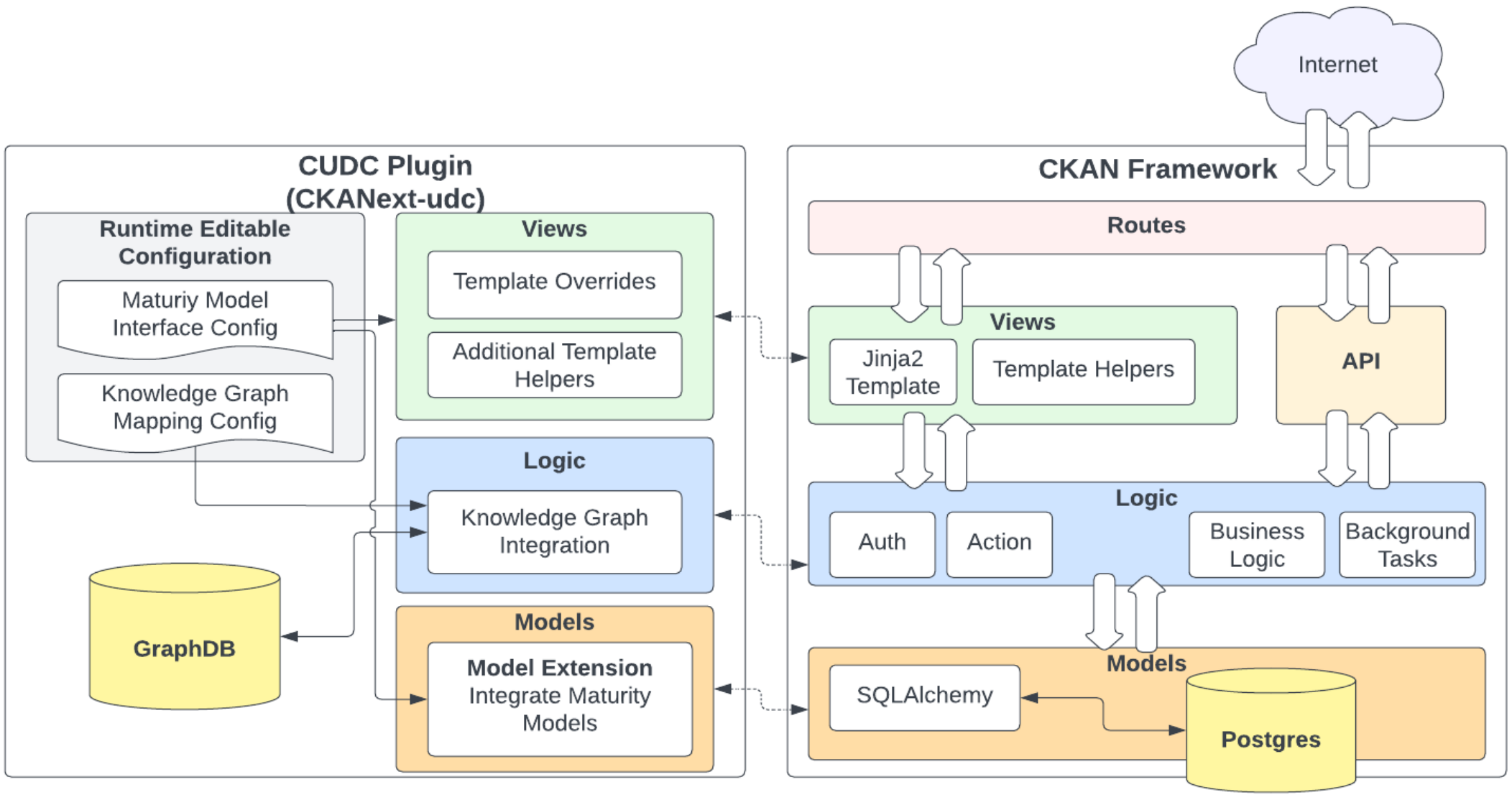

**Figure 5.** CUDC Plugin (CKANext-udc) Architecture.

The CKAN extension architecture, shown in **Figure 5**, is developed as a CKAN plugin. The plugin interacts with the CKAN architecture by modifying how catalogue metadata is collected from the user, how it is displayed to the user for data entry and viewing, and how data is stored in the database. Data entry views allow for grouping and entering maturity model properties seamlessly in CKAN.

First, the plugin refines the terminology by renaming "Dataset" to "Catalogue Entry" and "Resource" to "Dataset," a shift that is more in line with the maturity model's representation. Second, the maturity model is seamlessly integrated into both the edit and view pages of each catalogue entry. Third, the plugin introduces an advanced filter for improved search of the catalogue search. Beyond these template changes, the plugin adds more depth by overriding the default CKAN logic for managing data. For instance, when creating, updating, or deleting a catalogue entry, additional logic ensures that the knowledge graph is simultaneously updated based on the predefined mapping configuration between CKAN fields and the ontology. Lastly, the plugin enhances CKAN's foundational model by integrating the maturity model fields into CKAN's default dataset metadata model. The new plugin and CKAN's original model reside in the Postgres database. The plugin provides the ability to store catalogue data in a knowledge graph, as discussed below. By storing the metadata in both a knowledge graph and CKAN's internal database allows the plugin to be backward compatible with previous versions of CKAN, to version 2.11. The plug maintains all existing CKAN functionalities, including its API endpoint for importing and exporting catalogue data. Advanced users retain the ability to manage catalogue entries programmatically through CKAN's Python interface and API-based scripts.

---

[21] CKAN: https://docs.ckan.org/
[22] CKAN Plugin: https://github.com/csse-uoft/ckanext-udc
[23] Dataset Maturity Model Ontology: https://github.com/csse-uoft/maturity-model-ontology

## Maturity Model Views

A user can enter a new catalogue entry and its maturity levels properties, as shown in Figure 6. Each maturity level can be completed either partially or entirely. The plugin computes and displays the maturity levels' completion percentage for a catalogue entry. Maturity Model properties defined in the ontology are represented as text fields, date-time fields, or dropdown-select fields. Values in dropdown-select fields can either be entered manually or fetched from an ontology that defined the property, ensuring that the metadata aligns with imported ontologies. For example, the dataset file type in level 3 is represented as an instance of the class *dct:MediaType*. As such, it may include any of the types in IANA[24], including simple formats such as "json" and "csv", application formats such as "pdf" and "vnd.ms-excel", or complex formats, such as "ace+json", "csvm+json", "csv-schema", and so on. The formats are displayed to a user during the data entry step as one of the possible options.

[24] Internet Assigned Numbers Authority (IANA): https://www.iana.org/assignments/media-types/media-types.xhtml

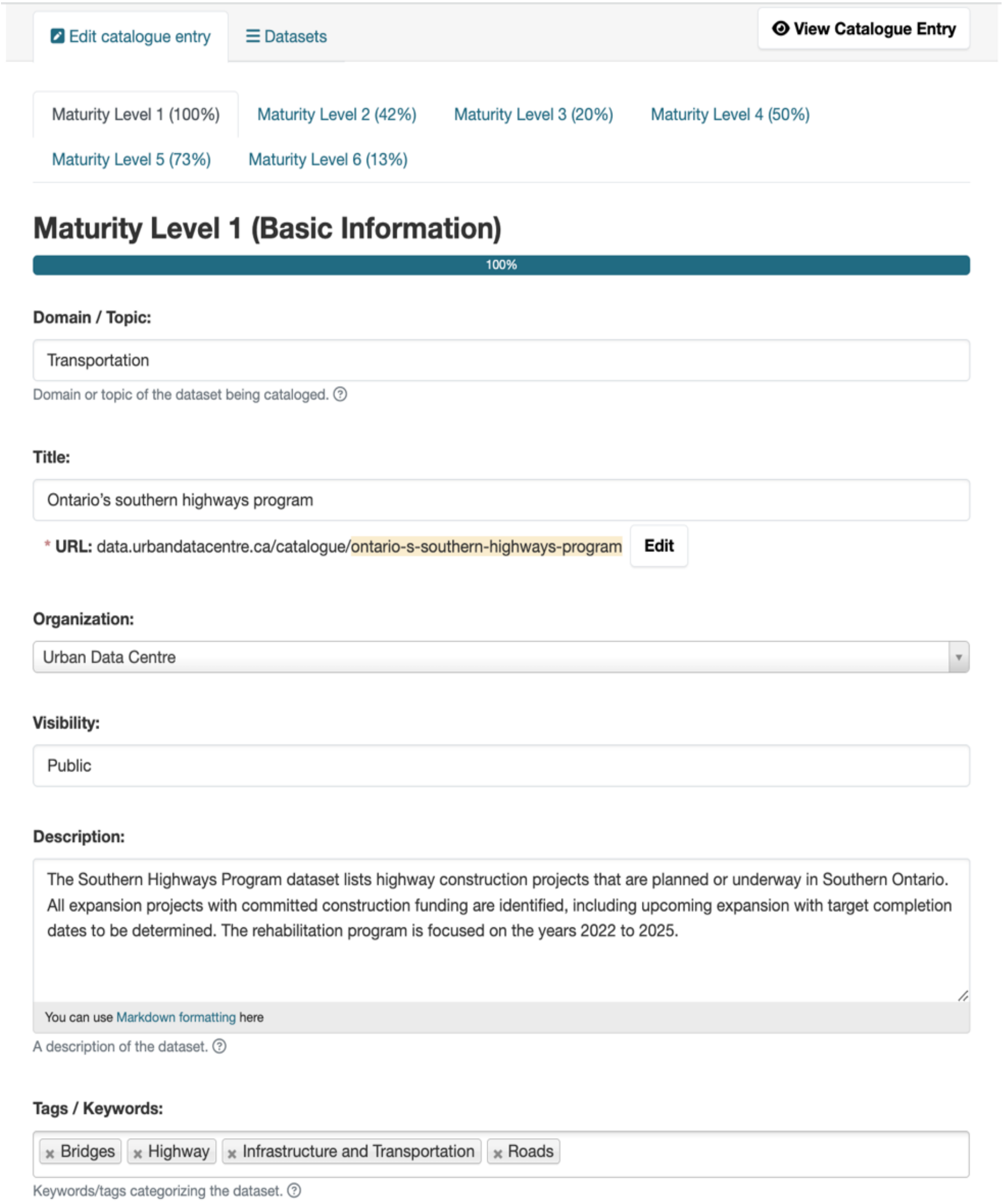

**Figure 6.** Maturity Model entry form, with five maturity levels and percentage of completed fields.

## Maturity Model View Configuration

Administrators have the flexibility to adjust the plugin settings directly from the web interface, including the data entry screen and mapping between entry fields and the maturity model properties. CKAN provides several predefined dataset properties, such as "title," "tags," "notes," "author," and more. The DMCMM used by the plugin is defined in a configuration file (JSON format), as illustrated in Figure 7, to organize the layout and properties when they are entered and displayed. Each maturity level tab, as shown in Figure 6, has a configuration entry under the "maturity_model" dictionary key. Each maturity level has a "title," "name," and an array of "fields" that specify the level's properties. Existing CKAN properties that are also specified in the maturity model are mapped to the model's ontology properties. If the property is a CKAN property, "ckanField" is used and the property name is the value in "ckanField". If the maturity model introduced a new property, the configuration requires a “name” and “label” entry in the “fields” key.

```
{
  "maturity_model": [{
      "title": "Maturity Level 1 (Basic Information)",
      "name": "maturity_level_1",
      "fields": [
        { // Maturity Model Property
          "name": "theme",
          "label": "Domain / Topic"
          ...
        },
        { // CKAN Property
          "ckanField": "title"
          ...
        }
        ...
  }]
}
```

**Figure 7.** CKAN Maturity Model configuration for data entry.

Users can search for catalogue entries using a text search field or a filter applicable to several key maturity model properties, as shown in Figure 8a. Search functionality utilizes CKAN’s built-in support for Apache Solr library[25] for indexing and searching text data. Results are shown in the view list screen. For catalogue entries that match a filter or search query, a side panel, shown in Figure 8b, displays an aggregate sum for indexed properties. The aggregate results are limited to the subset of results that match the original query.

By utilizing the text search and filter search, the plugin allows us to how users of the catalogue search for datasets. For example, free text search provides insights into how users refer to or spell various maturity model properties. Filter search provides us with similar monitoring capabilities but on a catalogue property level.

## Storing Dataset Metadata Capability Maturity Models in a Knowledge Graph

All catalogue entries, and their maturity model data, are stored in Postgres by CKAN. There is also an option to configure the CKAN plugin to connect with a graph database. The graph database uses the DMCMM ontology as its schema to store catalogue entries. To do this, one must supply the necessary mappings to the data present in the knowledge graph. As catalogue entries are updated or deleted, the plugin produces corresponding SPARQL queries based on these mappings, ensuring that the knowledge graph remains in sync with CKAN's database. An illustrative example of this mapping configuration utilizes a structure reminiscent of JSON-LD in Figure 9. For dynamic elements, such as generating an UUID value, the syntax allows one to provide a JavaScript helper function[26] call, such as “generate_uuid(),” as demonstrated below.

a) Filter Search Screen (entry screen) b) Filter Aggregation (side panel)

[25] Apache Solr library: https://solr.apache.org/
[26] The plugin allows for custom helper functions to be defined in “ckanext/udc/graph/mapping_helpers.py“

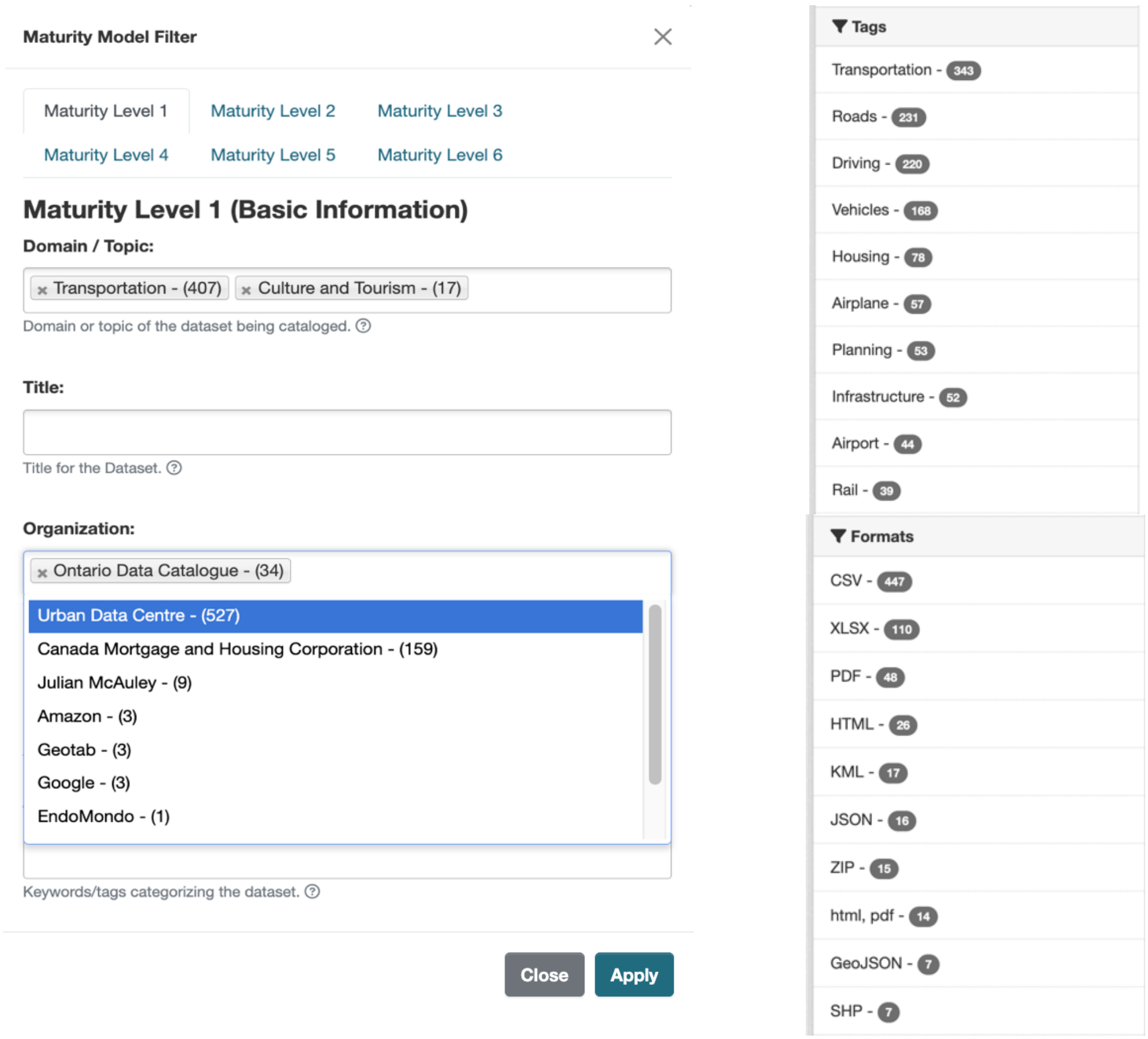

**Figure 8.** Maturity Model search capability.

```
{
  "mappings": {
    "@context": {
        // Define various namespaces that are used below.
        "xsd": "http://www.w3.org/2001/XMLSchema#",
        "dcat": "http://www.w3.org/ns/dcat#",
        "foaf": "http://xmlns.com/foaf/0.1/",
        "dct": "http://purl.org/dc/terms/"
    },
    // The URI of the catalogue entry, values in the curly bracket { }
    // will be evaluated at runtime. `ckanField` is a dictionary
    // and `ckanField.id` is a unique id of this catalogue entry.
    "@id": "http://data.urbandatacentre.ca/catalogue/{ckanField.id}",
    // The RDF Type of the catalogue
    "@type": "http://data.urbandatacentre.ca/Catalogue",
    // Author name and email are mapped into a `foaf:Agent` instance. Contents in the
    // curly bracket { } will be evaluated in the runtime.
    "dct:creator": {
      "@id": "http://data.urbandatacentre.ca/creator/{generate_uuid()}",
      "@type": "foaf:Agent",
      "foaf:mbox": "{ckanField.author_email}",
      "foaf:name": "{ckanField.author}"
    },
    // title is mapped to `dct:title`
    "dct:title": {
      "@type": "xsd:string",
      "@value": "{ckanField.title}"
    },
    // Published Date is mapped to `dct:issued`
    "dct:issued": {
      "@type": "xsd:date",
      "@value": "{to_date(published_date)}" // to_date(…) is a helper function
    }
  }
}
```

**Figure 9.** CKAN Maturity Model configuration for data entry.

# Metadata Availability Evaluation

Over a period of 12 months, a team of eight cataloguers scanned the web for Canadian sourced urban-related datasets, both open, closed, and through a web service, primarily for the themes of "Transportation," "Housing," "Bylaws," "Homelessness," and "Culture and Tourism"[27]. 1,162 datasets were catalogued of which 83% were in the five themes. Figure 10 shows a word cloud of keywords associated with each catalogue entry. The keywords have a strong correlation with the selected themes. For each dataset, the cataloguers extracted as much information as available to complete the metadata properties in maturity model.

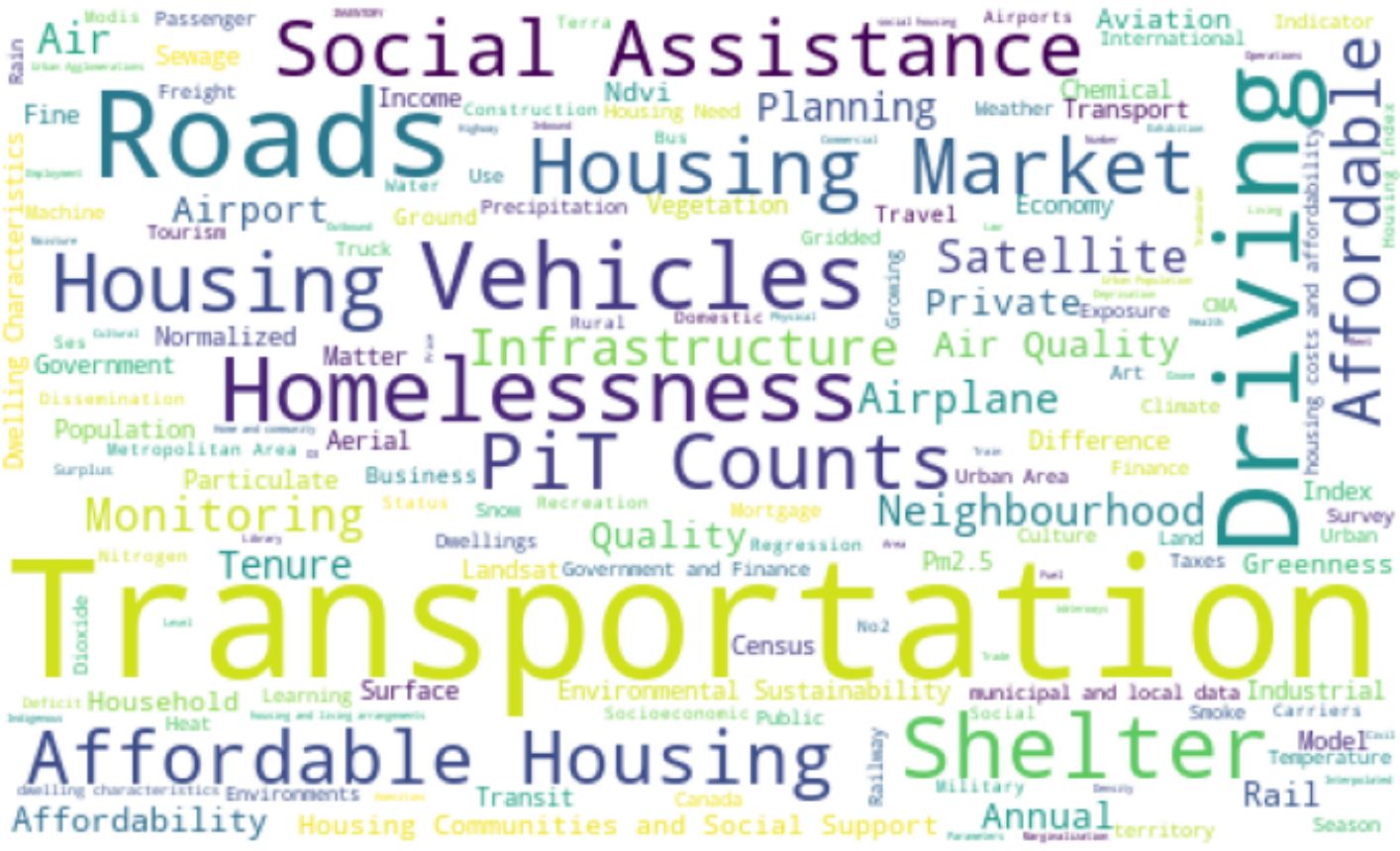

Figure 10. Word cloud of catalogue keywords.

The evaluation ascertains which dataset properties were most readily available, and rank maturity levels based on this availability. Towards this end, we first evaluate the completion rate of categories of information to give us a sense of what type of information is readily available to our cataloguers. Figure 11 shows the completion percentage for each of the seven categories of information. As was expected, in general, "Content" and "Access" related properties are most common, at above 60%. It was assumed that "Content" would have the most information but due to "file size" and "Metadata Identifier" not being available, "Content" has a low completed percentage. Information about the location and times in the "Temp-Geo", i.e. the temporal and geospatial coverage of the dataset, is available 60% of the time. Properties related to the "Ownership" of the dataset have a completion rate of 37%. While "Ownership" and "Access" may be closely related from a licensing perspective, accessibility was more readily available than contact ownership information. Hence, having access to the dataset may be more important than knowing the owner. This is reflected in the fact that "Access" properties are completed 33% more often than "Ownership" properties. Properties indicating the "Quality" of the dataset require some quality metric provided by the publisher. As can be seen, this information is rarely available at 24%. The completion rate of "Provenance" is also low, at 17%. This is due to the fact that provenance information is rarely provided, indicating that it may be considered as nonessential by dataset publishers. Finally, we note that "Statistics" category, indicating the size of the dataset, has the lowest completion rate at 12% with a relatively high standard deviation of 22, indicating this information is the most difficult to obtain.

[27] Prior to cataloguing Canadian urban datasets, two studies were undertaken to ascertain data requirements for research in the themes of Transportation (Pandya, 2023a) and Housing (Pandya, 2023b).

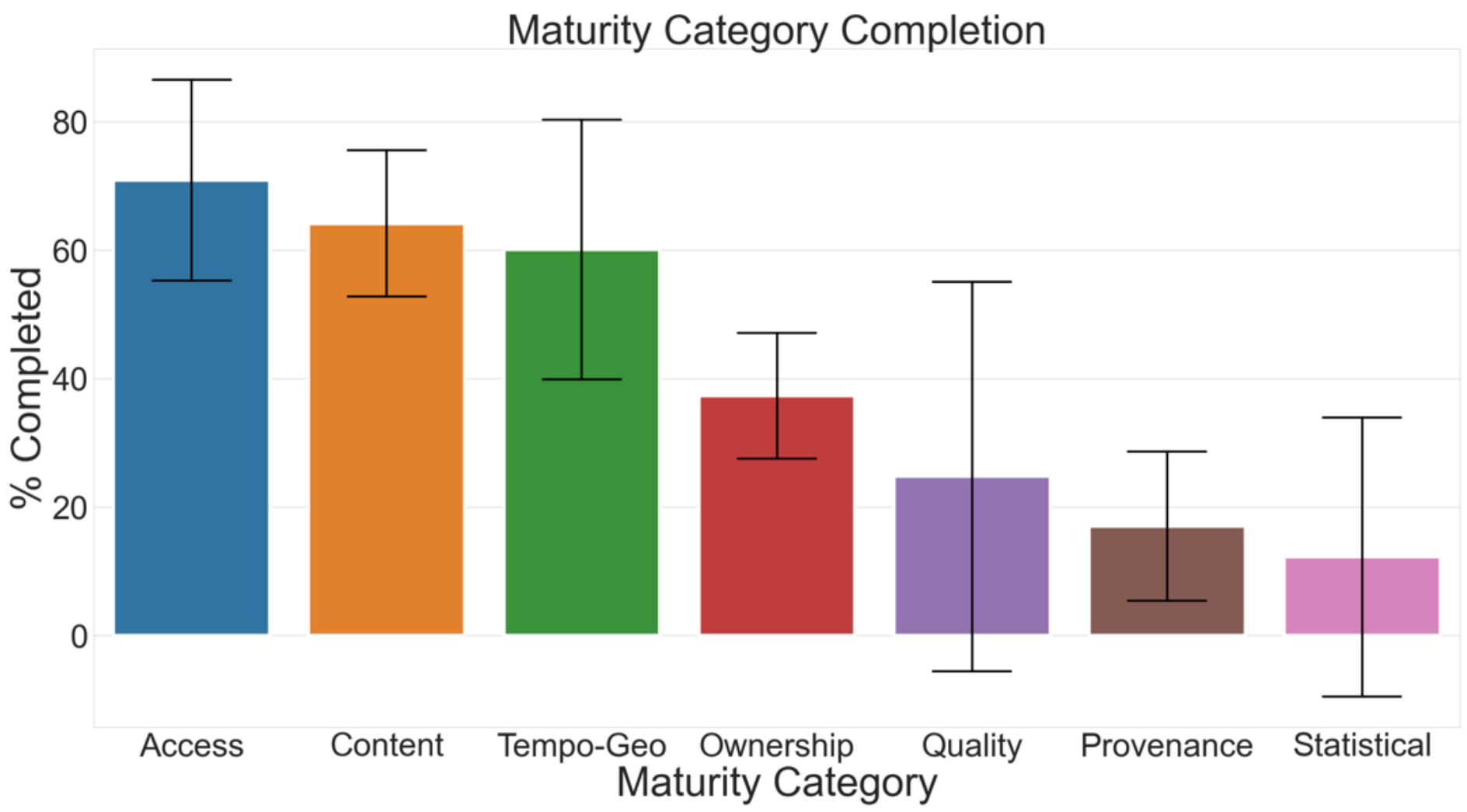

**Figure 11.** Percentage of Category Completion.

Finally, our evaluation focuses on ascertaining, for each level of the maturity model, the percentage of metadata properties that are available. The percentage of completed properties in each level is given in Figure 12. Maturity Level 1 emphasizes the metadata predominantly employed for dataset searches. The completion rate for this level stands at an average of 75%, accompanied by a standard deviation of 11. Maturity Level 2 focuses on access and ownership metadata. This level records a completion rate of 64% and a standard deviation of 12. Level 3 expands on content, provenance and temporal/geospatial information. It registered a 56% completion rate and a standard deviation of 17. Level 4 assesses the existence of data on individuals versus aggregates, any limits on use, and whether data pertinent to certain communities is captured or not. Level 4 has a completion rate of 32% an associated standard deviation of 14%. Maturity Level 5 focuses on attributes relevant to indigenous data management policies, having the lowest completion level at 0.5% and a high standard deviation of 4%. Finally, Maturity Level 6 is centred on the statistical and quality properties of the data, including the number of triples and concepts in triple stores or the row and column count in tabular datasets. The completion metric for this level is low at 15%, with a high standard deviation relative to the mean at 18.

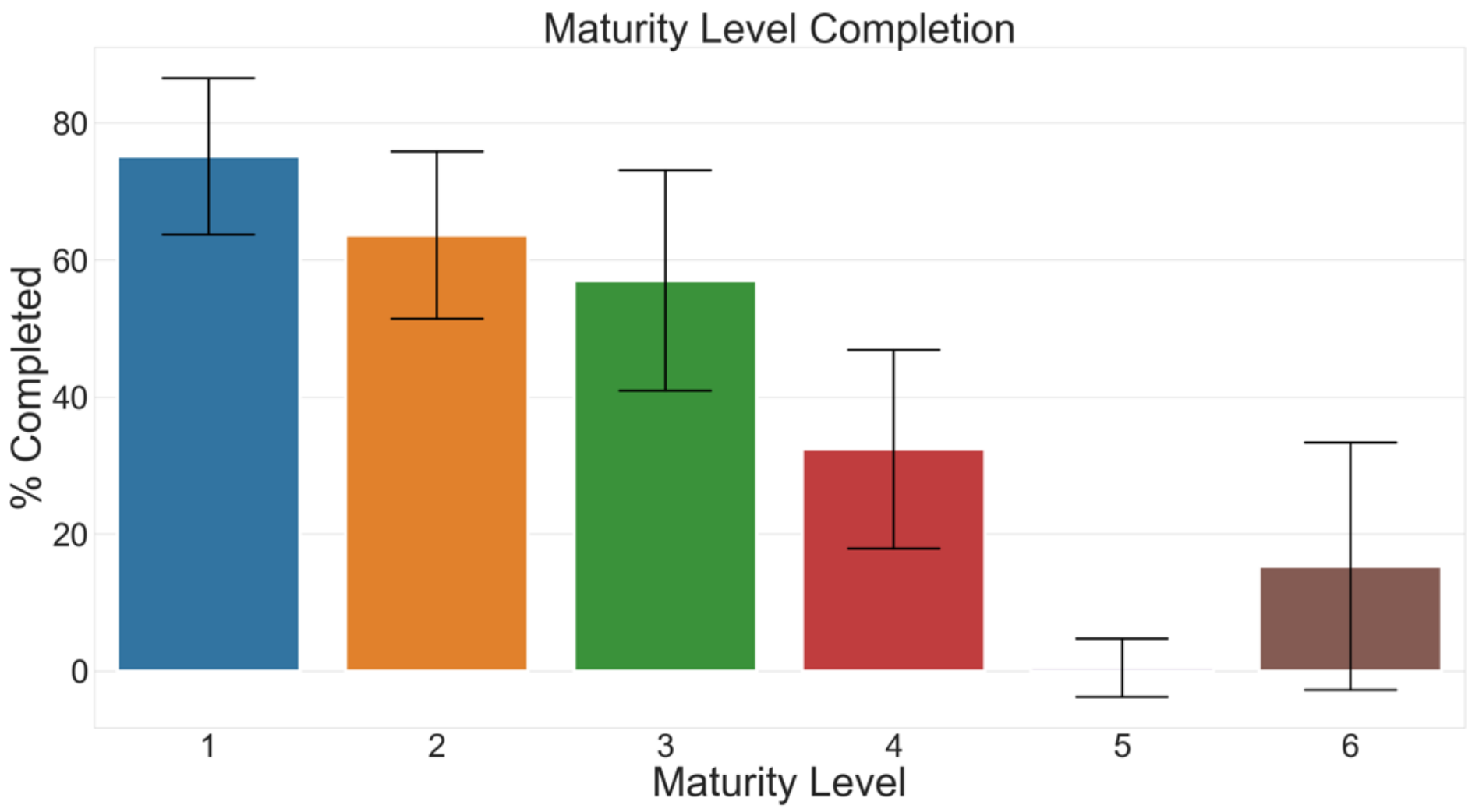

**Figure 12.** Percentage of Maturity Level Completion.

In accordance with our expectations, the availability of properties being readily available to cataloguers followed the maturity levels, with the exception of Maturity Level 5, as discussed below. It was also expected that Maturity Level 1 would be higher, above 80%. The low completion rate indicates missing information about file size and metadata found in the "Content" category of fields in level 1. We also note the high standard deviation of Levels 3, 4, 5, and 6, indicating that this availability varied greatly, and further analysis is needed. The low completion rate of Maturity Level 4 may be attributed to missing information about related datasets and versioning. After Maturity Level 5, level 6 manifested the most minimal completion rate. This can be attributed, initially, to the bifurcation of the statistical properties in Level 6 into two dataset categories: triple store and tabular. Thus, a completion rate of 100% would be contingent upon the existence of multimodal datasets encompassing both types. Furthermore, discerning this information mandates access to and enumeration of data points specific to their respective modes, a task that wasn't uniformly viable. Finally, the addition of "reviews" to the quality metric will increase the quality completion rate. This work is ongoing.

Based on our evaluation, we note that the basic descriptive properties of datasets in Level 1 are, in fact, the easiest to identify and can used as the initial search criteria for datasets. Descriptive properties at Level 2 are similarly obtainable to those in Level 3. This indicates that finding information about accessing the datasets requires as much effort as identifying the dataset's aggregation level. Level 4 properties about provenance, as well as other versions and related datasets was difficult to obtain by curators. Low level 5 completion rate indicates that more work is needed to clearly identify Indigenous data and policies related to data access and community ownership principles. Finally, the low completion rate of level 6 properties indicates that the dual nature of the properties requires a separation of modality categories, which are counted independently to better capture completion rates, and that description for the quality of the data is not readily available.

# Conclusion

This paper highlights the challenges of finding relevant data despite the abundance of available data. It discusses the obstacles in finding data, such as poor metadata, inappropriate data presentation, and difficulty locating the desired data. The paper also emphasizes the issue of dataset creators needing to gain knowledge about the appropriate metadata to specify for their datasets. The complexity of the task is exemplified by the extensive range of questions and properties specified in datasheets and the various vocabularies like DCAT, schema.org, PROV, and DQV. The survey of dataset search literature identified the types of information frequently used by searchers to discover relevant datasets.

To address the challenge of documenting datasets, this paper proposes a dataset metadata capability maturity model. This model proposes a sequence of levels of metadata specification capability. The goal is to strike a balance between the effort required to document a dataset and the provision of sufficient metadata for discovery, determining relevance, and understanding dataset content. The selection of attributes included in the DMCMM has been informed by a number of requirements, including: 1) what is needed to find relevant datasets; 2) how does one access a dataset and what restrictions exist; 3) what information is available to understand the content of the database; 4) how was the dataset derived; 5) what information is necessary to evaluate FAIRneess of a dataset; and 6) what information is necessary to support indigenous data sovereignty.

The stratification of the metadata attributes has been guided at the lowest level by information needed to find a relevant dataset, and at the next level by how to access the dataset. Subsequent levels focus on content, provenance, indigenous information, and data quality & statistics.

The Maturity Model has been incorporated into CKAN through a plugin named CKANext-udc. This plugin enhances the dataset cataloguing process by introducing custom fields categorized into the maturity levels and rearranging them on the user interface as needed.

Moreover, it integrates with a graph database, storing each catalogue entry's maturity model data as a knowledge graph underpinned by the Maturity Model ontology. The ontology is defined in OWL, and the plugin adjusts the CKAN system in various ways, from data collection to presentation and storage. Notably, the terminology is adjusted to match the maturity model, enabling advanced filtering for better search. The plugin also modifies default CKAN logic, ensuring that when catalogue entries change, the knowledge graph is updated accordingly. All conventional CKAN features, including the API endpoint for data import and export, are preserved, and advanced users can still manage entries using CKAN's Python interface and API scripts.

Finally, the Canadian Urban Data Catalogue (CUDC) has been implemented on the CKAN platform with the CKANext-udc plug-in (data.urbandatacentre.ca). It currently contains over 1,200 catalogued datasets.

# Appendix 1: Schema.org Dataset Properties

Dataset - Schema.org Type 2022-11-21, 7:45 PM

Schema.org Docs Schemas Validate About

## Dataset

*A Schema.org Type*

Thing > CreativeWork > Dataset

[more...]

A body of structured information describing some topic(s) of interest.

| Property | Expected Type | Description |
|---|---|---|
| **Properties from Dataset** | | |
| **distribution** | DataDownload | A downloadable form of this dataset, at a specific location, in a specific format. This property can be repeated if different variations are available. There is no expectation that different downloadable distributions must contain exactly equivalent information (see also DCAT on this point). Different distributions might include or exclude different subsets of the entire dataset, for example. |
| **includedInDataCatalog** | DataCatalog | A data catalog which contains this dataset. Supersedes catalog, includedDataCatalog.<br>Inverse property: dataset |
| **issn** | Text | The International Standard Serial Number (ISSN) that identifies this serial publication. You can repeat this property to identify different formats of, or the linking ISSN (ISSN-L) for, this serial publication. |
| **measurementTechnique** | Text or URL | A technique or technology used in a Dataset (or DataDownload, DataCatalog), corresponding to the method used for measuring the corresponding variable(s) (described using variableMeasured). This is oriented towards scientific and scholarly dataset publication but may have broader applicability; it is not intended as a full representation of measurement, but rather as a high level summary for dataset discovery.<br><br>For example, if variableMeasured is: molecule concentration, measurementTechnique could be: "mass spectrometry" or "nmr spectroscopy" or "colorimetry" or "immunofluorescence".<br><br>If the variableMeasured is "depression rating", the measurementTechnique could be "Zung Scale" or "HAM-D" or "Beck Depression Inventory".<br><br>If there are several variableMeasured properties recorded for some given data object, use a PropertyValue for each variableMeasured and attach the corresponding measurementTechnique. |
| **variableMeasured** | PropertyValue or Text | The variableMeasured property can indicate (repeated as necessary) the variables that are measured in some dataset, either described as text or as pairs of identifier and description using PropertyValue. |

https://schema.org/Dataset Page 1 of 11

**Figure 13.** Sample of schema.org Dataset Class Definition.

# Appendix 2: Datasheets for Datasets Questions

| Category | Question |
|---|---|
| Motivation | For what purpose was the dataset created? Was there a specific task in mind? Was there a specific gap that needed to be filled? |
| | Who created the dataset (for example, which team, research group) and on behalf of which entity (for example, company, institution, organization)? |
| | Who funded the creation of the dataset? |
| Composition | What do the instances that comprise the dataset represent (for example, documents, photos, people, countries)? Are there multiple types of instances (for example, movies, users, and ratings; people and interactions between them; nodes and edges)? |
| | How many instances are there in total (of each type, if appropriate)? |
| | Does the dataset contain all possible instances or is it a sample (not necessarily random) of instances from a larger set? If the dataset is a sample, then what is the larger set? Is the sample representative of the larger set (for example, geographic coverage)? If so, please describe how this representativeness was validated/ verified. If it is not representative of the larger set, please describe why not (for example, to cover a more diverse range of instances, because instances were withheld or unavailable) |
| | What data does each instance consist of? "Raw" data (for example, unprocessed text or images) or features? |
| | Is there a label or target associated with each instance? |
| | Is any information missing from individual instances? If so, please provide a description, explaining why this information is missing (for example, because it was unavailable). This does not include intentionally removed information, but might include, for example, redacted text. |
| | Are relationships between individual instances made explicit (for example, users' movie ratings, social network links)? If so, please describe how these relationships are made explicit. |
| | Are there recommended data splits (for example, training, development/validation, testing)? If so, please provide a description of these splits, explaining the rationale behind them. |
| | Are there any errors, sources of noise, or redundancies in the dataset? |
| | Is the dataset self-contained, or does it link to or otherwise rely on external resources (for example, websites, tweets, other datasets)? If it links to or relies on external resources, a) are there guarantees that they will exist, and remain constant, over time; b) are there official archival versions of the complete dataset (that is, including the external resources as they existed at the time the dataset was created); c) are there any restrictions (for example, licenses, fees) associated with any of the external resources that might apply to a dataset consumer? Please provide descriptions of all external resources and any |

| Category | Question |
|---|---|
| | restrictions associated with them, as well as links or other access points, as appropriate. |
| | Does the dataset contain data that might be considered confidential (for example, data that is protected by legal privilege or by doctor–patient confidentiality, data that includes the content of individuals' non-public communications)? |
| | Does the dataset contain data that, if viewed directly, might be offensive, insulting, threatening, or might otherwise cause anxiety? If so, please describe why. |
| | Does the dataset identify any subpopulations (for example, by age, gender)? If so, please describe how these subpopulations are identified and provide a description of their respective distributions within the dataset. |
| | .Is it possible to identify individuals (that is, one or more natural persons), either directly or indirectly (that is, in combination with other data) from the dataset? |
| | Does the dataset contain data that might be considered sensitive in any way (for example, data that reveals race or ethnic origins, sexual orientations, religious beliefs, political opinions or union memberships, or locations; financial or health data; biometric or genetic data; forms of government identification, such as social security numbers; criminal history)? |
| Collection Process | 21. How was the data associated with each instance acquired? Was the data directly observable (for example, raw text, movie ratings), reported by subjects (for example, survey responses), or indirectly inferred/ derived from other data (for example, part-of-speech tags, model-based guesses for age or language)? If the data was reported by subjects or indirectly inferred/derived from other data, was the data validated/verified? If so, please describe how. |
| | 22. What mechanisms or procedures were used to collect the data (for example, hardware apparatuses or sensors, manual human curation, software programs, software APIs)? How were these mechanisms or procedures validated? |
| | 23. If the dataset is a sample from a larger set, what was the sampling strategy (for example, deterministic, probabilistic with specific sampling probabilities)? |
| | 24. Who was involved in the data collection process (for example, students, crowdworkers, contractors) and how were they compensated (for example, how much were crowdworkers paid)? |
| | 25. Over what timeframe was the data collected? Does this timeframe match the creation timeframe of the data associated with the instances (for example, recent crawl of old news articles)? If not, please describe the timeframe in which the data associated with the instances was created. |
| | 26. Were any ethical review processes conducted (for example, by an institutional review board)? If so, please provide a description of these review processes, including the outcomes, as well as a link or other access point to any supporting documentation. |

| Category | Question |
|---|---|
| | If the dataset does not relate to people, you may skip the remaining questions in this section. |
| | 27. Did you collect the data from the individuals in question directly, or obtain it via third parties or other sources (for example, websites)? |
| | 28.Were the individuals in question notified about the data collection? If so, please describe (or show with screenshots or other information) how notice was provided, and provide a link or other access point to, or otherwise reproduce, the exact language of the notification itself. |
| | 29. Did the individuals in question consent to the collection and use of their data? If so, please describe (or show with screenshots or other information) how consent was requested and provided, and provide a link or other access point to, or otherwise reproduce, the exact language to which the individuals consented. |
| | 30. If consent was obtained, were the consenting individuals provided with a mechanism to revoke their consent in the future or for certain uses? If so, please provide a description, as well as a link or other access point to the mechanism (if appropriate). |
| | 31.Has an analysis of the potential impact of the dataset and its use on data subjects (for example, a data protection impact analysis) been conducted? If so, please provide a description of this analysis, including the outcomes, as well as a link or other access point to any supporting documentation. |
| Preprocessing/ Cleaning/ Labeling | 33. Was any preprocessing/cleaning/labeling of the data done (for example, discretization or bucketing, tokenization, part-of-speech tagging, SIFT feature extraction, removal of instances, processing of missing values)? If so, please provide a description. If not, you may skip the remaining questions in this section. |
| | 34. Was the “raw” data saved in addition to the preprocessed/cleaned/ labeled data (for example, to support unanticipated future uses)? If so, please provide a link or other access point to the “raw” data. |
| | 35. Is the software that was used to preprocess/clean/label the data available? If so, please provide a link or other access point. |
| Uses | 37. Has the dataset been used for any tasks already? If so, please provide a description. |
| | 38. Is there a repository that links to any or all papers or systems that use the dataset? If so, please provide a link or other access point. |
| | 39. What (other) tasks could the dataset be used for? |
| | 40. Is there anything about the composition of the dataset or the way it was collected and preprocessed/ cleaned/labeled that might impact future uses? For example, is there anything that a dataset consumer might need to know to avoid uses that could result in unfair treatment of individuals or groups (for example, stereotyping, quality of service issues) or other risks or harms (for |

| Category | Question |
|---|---|
| | example, legal risks, financial harms)? If so, please provide a description. Is there anything a dataset consumer could do to mitigate these risks or harms? |
| | 41. Are there tasks for which the dataset should not be used? If so, please provide a description. |
| Distribution | 43. Will the dataset be distributed to third parties outside of the entity (for example, company, institution, organization) on behalf of which the dataset was created? If so, please provide a description. |
| | 44. How will the dataset be distributed (for example, tarball on website, API, GitHub)? Does the dataset have a digital object identifier (DOI)? |
| | 45. When will the dataset be distributed? |
| | 46. Will the dataset be distributed under a copyright or other intellectual property (IP) license, and/or under applicable terms of use (ToU)? If so, please describe this license and/ or ToU, and provide a link or other access point to, or otherwise reproduce, any relevant licensing terms or ToU, as well as any fees associated with these restrictions. |
| | 47. Have any third parties imposed IP-based or other restrictions on the data associated with the instances? If so, please describe these restrictions, and provide a link or other access point to, or otherwise reproduce, any relevant licensing terms, as well as any fees associated with these restrictions. |
| | 48. Do any export controls or other regulatory restrictions apply to the dataset or to individual instances? If so, please describe these restrictions, and provide a link or other access point to, or otherwise reproduce, any supporting documentation. |
| Maintenance | 50. Who will be supporting/hosting/maintaining the dataset? |
| | 51. How can the owner/curator/ manager of the dataset be contacted (for example, email address)? |
| | 52. Is there an erratum? If so, please provide a link or other access point. |
| | 53. Will the dataset be updated (for example, to correct labeling errors, add new instances, delete instances)? If so, please describe how often, by |
| | whom, and how updates will be communicated to dataset consumers (for example, mailing list, GitHub)? |
| | 54. If the dataset relates to people, are there applicable limits on the retention of the data associated with the instances (for example, were the individuals in question told that their data would be retained for a fixed period of time and then deleted)? If so, please describe these limits and explain how they will be enforced. |
| | 55. Will older versions of the dataset continue to be supported/hosted/ maintained? If so, please describe how. If not, please describe how its obsolescence will be communicated to dataset consumers. |

# Appendix 2: FAIR Principles Properties

The FAIR principles (Findable, Accessible, Interoperable, Retrievable) can be inferred from the DMCMM maturity levels. In **Table 29**, we propose key data-related principles and their properties. The properties are defined with a boolean range, indicating whether the FAIR principle was met or not. Additional validation for compatibility with specific technologies, formats, or methods require additional steps by the user.

**Table 29.** FAIR Priniciples Properties.

| Category | Description | Property | Value Restriction |
|---|---|---|---|
| Content | Data complies with a community standard | fair:hasRDA_R1_3_01D | xsd:boolean |
| | Data uses knowledge representation expressed in standardised format | fair:hasRDA_I1_01D | xsd:boolean |
| | Data uses machine-understandable knowledge representation | fair:hasRDA_I1_02D | xsd:boolean |
| | Data uses FAIR-compliant vocabularies | fair:hasRDA_I2_01D | xsd:boolean |
| | Data includes references to other data | fair:hasRDA_I3_01D | xsd:boolean |
| Access | Data is accessible through an access protocol that supports authentication and authorisation | fair:hasRDA_A1_2_01D | xsd:boolean |
| | Data can be accessed manually (i.e., with human intervention) | fair:hasRDA_A1_02D | xsd:boolean |
| | Data identifier resolves to a digital object | fair:hasRDA_A1_03D | xsd:boolean |
| | Data is accessible through standardised protocol | fair:hasRDA_A1_04D | xsd:boolean |
| | Data can be accessed automatically (i.e. by a computer program) | fair:hasRDA_A1_05D | xsd:boolean |
| | Data is accessible through a free access protocol | fair:hasRDA_A1_1_01D | xsd:boolean |

# Appendix 3: Indigenous-related Properties

Datasets about Indigenous communities and OCAP properties. In **Table 30**, we propose a set of properties to capture information needed to identify communities and stewards of Indigenous datasets.

**Table 30.** Dataset Maturity Level 4 Definition.

| Category | Description | Property | Value Restriction |
|---|---|---|---|
| Content | Contains Indigenous Data | containsIndigenousData | xsd:boolean |
| Ownership | Indigenous community permission (who gave permission) | indigenousRightsHolder | foaf:IndigenousCommunityGroup |
| Temporal/ Geospatial | Indigenous communities from which data is derived | spatialIndigenousCommunity (sub-property of dct:spatial) | dct:Location |

Notes:

- **containsIndividualData**: A boolean value indicating whether the data holds individualized data? If yes, the dataset is aggregated at the individual level.
- **containsIdentifiableIndividualData**: Does the data hold identifiable individual data that can be used to uniquely identify an individual data was collected about? If yes, the dataset is not anonymized.
- **containsIndigenousData**: Does the data hold data about Indigenous communities? If yes, the dataset should comply with OCAP principles (Ownership, Control, Access, Possession). See https://fnigc.ca/ocap-training/ for more information.

**Table 31.** IndigenousAgent Class Definition.

| ClassProperty | Property | Value Restriction |
|---|---|---|
| IndigenousCommunityGroup | rdfs:subClassOf | :CommunityGroup |

- **IndigenousCommunityGroup**: A subclass of :CommunityGroup that is made up of Indigenous members.

- **Table 32.** indigenousRightsHolder Property Definition.

| ClassProperty | Property | Value Restriction |
|---|---|---|
| indigenousRightsHolder | rdfs:subPropertyOf | dct:rightsHolder |

- **indigenousRightsHolder**: An agent that has the rights to manage access rights to indigenous data. That person can be Indigenous themselves or a non-Indigenous agent that acts as the steward for access rights to the data.

**Table 33.** spatialIndigenousCommunity Property Definition.

| ClassProperty | Property | Value Restriction |
|---|---|---|
| spatialIndigenousCommunity | rdfs:subPropertyOf | spatialCommunity |

- **spatialIndigenousCommunity**: A geospatial area occupied by or representative of an Indigenous community.